\documentclass[final]{arxiv}

\usepackage{graphicx}
\usepackage{newtxtext}
\usepackage{newtxmath}
\usepackage{natbib}
\usepackage{hyperref}
\hypersetup{
    colorlinks = true,
    urlcolor   = blue,
    citecolor  = black,
}
\usepackage{tikz} 
\DeclareRobustCommand\full  {\tikz[baseline=-0.6ex]\draw[thick] (0,0)--(0.5,0);}
\DeclareRobustCommand\dashed{\tikz[baseline=-0.6ex]\draw[thick,dashed] (0,0)--(0.54,0);}

\newcommand{\RomanNumeralCaps}[1]
\linenumbers

\title{Validation and extension of an analytic momentum availability model for the two-scale momentum theory of wind farm flows}

\author{Mads Baungaard\aff{1}
  \corresp{\email{mads.baungaard@eng.ox.ac.uk}}, Takafumi Nishino\aff{1} and Andrew Kirby\aff{2}}

\affiliation{\aff{1}Department of Engineering Science, Univeristy of Oxford, Parks Road, Oxford OX1 3PJ, UK
\aff{2}5 Coal Lane, London SW9 8GG, UK}

\begin{document}
\maketitle

\begin{abstract}

A key parameter in the two-scale momentum theory of wind farm flows is the momentum availability, which quantifies the supply of momentum to a wind farm from various different momentum transport mechanisms (advection, pressure gradient, Coriolis, turbulence and unsteadiness). In this study, the contribution of each of these mechanisms to the momentum availability is evaluated directly from large-eddy simulation (LES) data in order to validate an analytic momentum availability model (Kirby, Dunstan, \& Nishino, \textit{J. Fluid Mech.}, vol. 976, 2023, A24). Application of the model to six wind farm cases, three with different atmospheric boundary-layer (ABL) heights and three with different turbine layouts, shows that the full model performs well across all cases, but that its linearized version increasingly overpredicts the momentum availability for increasing ABL heights. It is found that the overprediction is related to the ABL Rossby number, and based on this observation, we propose an extension of the original linear model, which improves its accuracy for the considered cases and makes it more generally applicable, in particular to cases with tall ABL heights or strong Coriolis forcing.

\end{abstract}



\section{Introduction}
\label{sec:intro}

The last two decades have seen the emergence of many offshore wind farms, which each provides a large amount of green electricity and contributes to a more sustainable future. It is estimated that the whole electricity demand of Europe could be fulfilled with offshore wind farms covering an area of less than one-third of the North Sea \citep{sorensen_assessment_2023}. To aid such an ambitious development, theoretical understanding and modelling of wind farm aerodynamics are crucial, but the multi-scale nature of the flow makes it a challenging task \citep{Porte-Agel2020}.

The classic approach for modelling wind farm aerodynamics is to model the velocity deficit downstream of each wind turbine, i.e. the `wind turbine wake', with an analytical expression \citep[e.g.][]{jensen_1983_NOJ,bastankhah2014} and superpose the wakes to obtain the combined wind farm flow \citep[e.g.][]{katic1986}. Both the wake and superposition models are based on various simplifications to obtain simple analytical forms and do not take into account the response of the atmospheric boundary-layer (ABL) to the wind farm. Another classic approach is the `top-down models' \citep[e.g.][]{frandsen1992,Calaf2010}, which are for very large wind farms, where the flow can be treated as horizontally homogeneous in the interior of the farm and thus enables simple modelling of the ABL--wind farm interaction, although without the explicit effect of the turbine layout. A combination of the two above approaches was suggested by \citet{Stevens2016coupled}, which thereby includes both the effect of turbine layout and the ABL response, but also inherits the empirical assumptions and limitations of the constituent models.

An alternative approach known as the `two-scale momentum theory' was proposed by \citet{Nishino2020}. It is based on the conservation of momentum within the farm region and splits the multi-scale flow problem into internal (turbine/array-scale) and external (farm/ABL-scale) sub-problems. The internal sub-problem is mainly about the turbine-wake interactions, while the external sub-problem concerns the farm-scale processes which supply momentum to the farm, and the farm's interaction with the ABL. By solving the two sub-problems together, an estimate for the upper limit of the wind farm performance can be obtained. Furthermore, the power losses in the farm can be classified into turbine-scale (due to wakes impinging downstream turbines) and farm-scale (from the `farm-average' wind speed due to its interaction with the ABL) losses  \citep{Kirby2022}. This classification has later been extended to also include losses due to non-ideal turbine design \citep{nishino_power_2025}. \citet{Kirby2022} conducted a suite of 50 LESs of offshore wind farms with varying turbine layouts, which suggested that the farm-scale loss is typically more than twice as large as the turbine-scale loss. This has the important implication that there is a limited potential in turbine layout optimization and that there should be an increased focus on methods that increase wind entrainment into the farm as highlighted by \citet{stevens_understanding_2023}.

The two-scale momentum theory has been derived in a general manner, hence the choice of the models for the internal and external sub-problems can be varied. In this paper, we focus on the external sub-problem, which is characterized by the momentum availability factor, $M$, defined as the ratio of momentum transfer into the farm region with and without the turbines present, respectively. The simplest possible model for this quantity is obtained by assuming that there is no interaction between the farm and the ABL, in which case the momentum transfer is equal for the two situations, and therefore $M = 1$. This simple model was used in the initial study of \citet{nishino2016} and by \citet{west2020}. More realistically, when the turbines are present, they create a resistance to the ABL, which may enhance the various momentum transport mechanisms (advection, pressure gradient, Coriolis, turbulence and unsteady effects). Hence, it was hypothesized by \citet{Nishino2020} that the momentum availability should increase approximately linearly as a function of the farm-wind-speed reduction. The proportionality factor of the linear increase of $M$ was termed the `momentum response factor' (in later works, it is also called the `wind extractability factor'), $\zeta$. \citet{Patel2021} calculated $\zeta$ for a large wind farm in the North Sea using a numerical weather prediction (NWP) simulation and observed to vary significantly during the day. Using a similar simulation setup \citet{Kirby2022} found that $\zeta$ is also heavily dependent on the farm size, and later LESs confirmed this and showed that it is much less sensitive to the exact turbine layout \citep{kirby_turbine-_2025}. 

A more elaborate model of the momentum availability was developed by \citet{kirby_ana2023} through considering each of the momentum transport mechanisms mentioned earlier and modelling each of these independently. By combining the sub-models, with some approximations, an analytic model was obtained. A simplified model was also derived from the analytic model using some additional linearizations. Although it was shown that the models appear to perform well, the validity and generality of some underlying assumptions and linearizations could not be confirmed directly at that time as limited LES data was available. In this study, we perform an in-depth validation of each contributing mechanism, which is possible with the full 3D flow fields from the wind farm LES database by \citet{lanzilao2025}, and with the knowledge obtained from this in-depth validation we propose a new momentum availability model. 

An overview of the two-scale momentum theory and the current momentum availability models is given in § \ref{sec:theory}. The various momentum availability models are then tested against LES data in § \ref{sec:test_of_M}. In § \ref{sec:LES}, each momentum availability contribution is evaluated one by one and the validity of the various sub-model assumptions are quantified. A new momentum availability model is proposed in § \ref{sec:rossby}, which is based on the ABL Rossby number and can be seen as an extension of the earlier model from \citet{kirby_ana2023}. Finally, conclusions are given in § \ref{sec:conclusions}.

\section{Theory}
\label{sec:theory}

\subsection{Two-scale momentum theory}
The central equation of the two-scale momentum theory \citep{Nishino2020} is the non-dimensional farm momentum (NDFM) equation
\begin{equation}
    C_{T}^* \frac{\lambda}{C_{f0}} \beta^2 + \beta^{\gamma} = M ,
    \label{eq:NDFM}
\end{equation}
where $C_T^*$ is the internal farm thrust coefficient, $\lambda$ is the array density, $C_{f0}$ is the surface friction coefficient of the ABL, $\beta$ is the wind-speed reduction factor, $\gamma$ is the friction exponent, and $M$ is the momentum availability factor. The exact definitions of these parameters will be given in the summary of the NDFM derivation described below. A typical usage of the equation is to solve for $\beta$, given information or models of the remaining parameters, which can then be used to calculate the farm power efficiency.

The NDFM equation is derived through a control volume (CV) analysis of the streamwise mean momentum equation
\begin{equation}
    \frac{\partial \rho U_1}{\partial t} = -\frac{\partial \rho U_1 U_j}{\partial x_j} -\frac{\partial P}{\partial x_1} + \rho f_c U_2 + \frac{\partial \tau_{1j}}{\partial x_j} + f_1 ,
    \label{eq:Ueq}
\end{equation}
for a volume, $V_{\rm cv} = L W H_{\rm cv} = S_{\rm cv} H_{\rm cv}$, containing the wind farm. In the original derivation by \citet{Nishino2020}, the height of the CV was chosen to be similar to the boundary-layer height, $H_{\rm cv} \approx \delta_{\rm ABL}$, whereas in later works \citep[e.g.][]{kirby_ana2023} it has been chosen as $H_{\rm cv} = H_F$ (where typically $H_F = 2.5 z_{h}$ is chosen with $z_h$ being the turbine hub-height) to model the effects of turbulent entrainment above the wind farm more explicitly than implicitly. For rectangular farm layouts (both aligned and staggered), the length and width of the CV are typically chosen as the number of turbines times inter-spacing, i.e. $L = N_x s_x$ and $W = N_y s_y$, where $N_i$ and $s_i$ are the number of turbines and inter-spacing in the $i$'th direction. The NDFM equation will however be valid for any choice of CV as long as it encloses the wind farm. 

Denoting the volume-average over the CV with square brackets (i.e. $\left[ \cdot \right] \equiv V_{\rm cv}^{-1} \int_{V_{\rm cv}} \cdot dV$) and using Gauss's theorem for the advection and turbulent stress terms gives
\begin{align}
    \left[\frac{ \partial \rho U_1}{\partial t} \right]  &= -\frac{1}{V_{\rm cv}}\int_{\Omega_{\rm cv}}  \rho U_1 U_j dA_j -  \left[ \frac{\partial P}{\partial x_1}  - \rho f_c U_2   \right]   + \frac{1}{V_{\rm cv}} \int_{\Omega_{\rm cv}  \setminus  \Omega_w}  \tau_{1j} dA_j - \frac{S_{\rm cv}  \tau_w }{V_{\rm cv}} - \frac{T}{V_{\rm cv}},
    \label{eq:U_volavg}
\end{align}
where $\Omega_{\rm cv}$ is the surface of the CV, $\tau_{w} \equiv S_{\rm cv}^{-1} \int_{\Omega_w} \tau_{13} dA$ is the surface-averaged streamwise component of the wall shear stress, and $T \equiv - \int_{V_{\rm cv}} f_1 dV$ is the integrated body force stemming from the total turbine thrust forces. Multiplying (\ref{eq:U_volavg}) through with the CV volume and rearranging gives
\begin{align}
     T + S_{\rm cv}  \tau_{w}   &= \underbrace{-\int_{\Omega_{\rm cv}}  \rho U_1 U_j dA_j}_{X_{\rm adv}} \underbrace{- V_{\rm cv}  \left[ \frac{\partial P}{\partial x_1} \right]}_{X_{\rm PGF}}  + \underbrace{  V_{\rm cv} \left[\rho f_c U_2   \right]}_{X_{\rm Cor}}  +  \underbrace{\int_{\Omega_{\rm cv}  \setminus  \Omega_w}  \tau_{1j} dA_j}_{X_{\rm turb}} \underbrace{- V_{\rm cv} \left[\frac{ \partial \rho U_1}{\partial t} \right]}_{X_{\rm uns}},
\end{align}
where the two terms on the left-hand side are the momentum sinks from the turbine and surface friction drag, respectively, and the terms on the right-hand side are the momentum balancing terms from advection, pressure-gradient forcing, Coriolis, turbulence, and unsteadiness. A similar CV analysis can be made for a case without turbines (which has $T = 0$ and is denoted with a zero subscript), and this can be used to normalize the above equation, i.e.
\begin{align}
     \frac{T + S_{\rm cv}  \tau_{w}}{S_{\rm cv}  \tau_{w0}}   &= \frac{X_{\rm adv} + X_{\rm PGF}  + X_{\rm Cor}  +  X_{\rm turb} + X_{\rm uns}}{X_{\rm adv 0} + X_{\rm PGF 0}  + X_{\rm Cor 0}  +  X_{\rm turb 0} + X_{\rm uns 0}}.
     \label{eq:derived_NDFM}
\end{align}
The left-hand side can be rewritten to
\begin{equation}
    \frac{T + S_{\rm cv}  \tau_{w} }{S_{\rm cv}  \tau_{w0} } = C_{T}^* \frac{\lambda}{C_{f0}} \beta^2 + \beta^{\gamma} ,
    \label{eq:NDFM_LHS}
\end{equation}
from the definitions $C_T^* \equiv T/(\frac{1}{2} \rho  U_F^2 N_t A_d)$, $U_F \equiv [U_1]$, $\lambda \equiv N_t A_d/S_{\rm cv}$, $C_{f0} \equiv  \tau_{w0} / (\frac{1}{2} \rho U_{F0}^2)$, $\beta \equiv U_F/U_{F0}$, and $\gamma = \textrm{log}_{\beta} \left( \frac{\tau_{w}}{\tau_{w0}} \right)$, hence (\ref{eq:derived_NDFM}) is, in fact, the NDFM equation, and the momentum availability factor is
\begin{equation}
    M \equiv \frac{X_F}{X_{F0}} = \frac{X_{\rm adv} + X_{\rm PGF}  + X_{\rm Cor}  +  X_{\rm turb} + X_{\rm uns}}{X_{\rm adv 0} + X_{\rm PGF 0}  + X_{\rm Cor 0}  +  X_{\rm turb 0} + X_{\rm uns 0}} .
    \label{eq:fullM}
\end{equation}

The contributions of the various physical processes to $M$ can conveniently be split into separate parts by rewriting (\ref{eq:fullM}) as
\begin{align}
    M &= 1 + \frac{X_F - X_{F0}}{X_{F0}} \label{eq:M_contributions}\\ &= 1 + \underbrace{ \frac{X_{\rm adv} - X_{\rm adv 0}}{X_{F0}}}_{\Delta M_{\rm adv}} + \underbrace{ \frac{X_{\rm PGF} - X_{\rm PGF 0}}{X_{F0}}}_{\Delta M_{\rm PGF}}
+ \underbrace{ \frac{X_{\rm Cor} - X_{\rm Cor 0}}{X_{F0}}}_{\Delta M_{\rm Cor}}
+ \underbrace{ \frac{X_{\rm turb} - X_{\rm turb 0}}{X_{F0}}}_{\Delta M_{\rm turb}}
+ \underbrace{ \frac{X_{\rm uns} - X_{\rm uns 0}}{X_{F0}}}_{\Delta M_{\rm uns}} , \nonumber
\end{align}
where one can note that $X_{F0} = S_{\rm cv} \tau_{w0}$. To solve the NDFM equation, one must model $M$, or each individual sub-part of it.

\subsection{Analytical momentum availability models}
The simplest model for $M$ is to assume that all $\Delta M$ contributions in (\ref{eq:M_contributions}) are negligibly small, hence
\begin{equation}
    M_{\rm constant} = 1 ,
\end{equation}
which corresponds to assuming that the momentum available to the wind farm is fixed, i.e. there is no ABL response to the farm. This is a too simplistic model for an infinitely large wind farm and not applicable to finite-size wind farms, as will be shown with the LES data in § \ref{sec:test_of_M}. 

A linear model for $M$ was proposed by \citet{Nishino2020} 
\begin{equation}
    M_{\rm lin} = 1 + \zeta (1-\beta) , \label{eq:M_linear}
\end{equation}
and further studied by \citet{Patel2021} through numerical weather prediction (NWP) simulations with a wind farm modelled as an increased surface roughness patch. They concluded that the linear relationship appeared to satisfactorily fit their data with a momentum response parameter $\zeta \sim \mathcal{O}(10^1)$ depending on the atmospheric conditions. Using a similar methodology, \citet{Kirby2022} noted that $\zeta$ strongly depends on the wind farm size with smaller $\zeta$ values being observed for larger wind farms.

A more elaborate model of $M$ was developed by \citet{kirby_ana2023} who considered each term of $M$ in (\ref{eq:M_contributions}) and used various physical arguments to construct the following analytical sub-models
\begin{align}
    \Delta M_{\rm{adv}} &= \frac{ H_F}{L C_{f0}}  \left( \beta_{\rm local}(0)^2 - \beta_{\rm local}(L)^2  \right) , \label{eq:Madv_model}\\
    \Delta M_{\rm{PGF}} &=    2  \frac{H_F}{L C_{f0}}   \left(1 - \beta_{\rm local}(0)^2 \right), \label{eq:Mpgf_model}\\
    \Delta M_{\rm{Cor}} &=    0, \label{eq:Mcor_model} \\
    \Delta M_{\rm{turb}} 
    &=   M + M\left(\frac{\tau_{t0}}{\tau_{w0}} - 1\right) \frac{h_0}{h}  - \frac{  \tau_{t0}}{\tau_{w0}}, \label{eq:Mstr_model} \\
    \Delta M_{\rm{uns}} 
    &=   0. \label{eq:Muns_model}
\end{align}
Here $\beta_{\rm local}(x) \equiv (H_F W U_{F0})^{-1} \int_0^{H_F} \int_0^W U_1 dy dz$ is the local wind-speed reduction factor, $h$ is the average ABL height over the farm region (which is pushed upwards due to the presence of the farm), and $\tau_{t} \equiv S_{\rm cv}^{-1} \int_{\Omega_t} \tau_{13} dA$ is the surface-averaged streamwise shear stress at the top of the CV. The 0 subscript again denotes the no-turbine scenario. It was then further assumed that $1 + \beta^2    = \beta_{\rm local}(0)^2 + \beta_{\rm local}(L)^2$, which we refer to in this work as the `advection-pressure (AP) approximation', as it conveniently can be utilized to combine and simplify the advection and pressure-gradient forcing (PGF) terms to
\begin{align}
     \Delta M_{\textrm{adv}} + \Delta M_{\textrm{PGF}} &= \frac{H_F}{L C_{f0}}  \left( 1 - \beta^2   \right) . \label{eq:Madv_and_Mpgf_sa}
\end{align}
\citet{kirby_ana2023} also derived an approximate model for the ABL height displacement 
\begin{equation}
    \frac{h_0}{h} = \beta ,
    \label{eq:bl_height_ratio_model}
\end{equation}
which is used in the stress term model (\ref{eq:Mstr_model}). The final combined model is thus
\begin{equation}
    M_{\rm KDN1} = \frac{1 + \frac{H_F}{L C_{f0}}  \left( 1 - \beta^2   \right)     - \frac{  \tau_{t0}}{\tau_{w0}}}{\beta \left(1 -\frac{\tau_{t0}}{\tau_{w0}}\right)} .\label{eq:M_KDN1}
\end{equation}
This momentum availability model was successfully used in conjunction with the NDFM equation to predict the wind farm efficiency for different wind farm LES cases by \citet{kirby_ana2023}. The accuracy and assumptions of the sub-models leading to (\ref{eq:M_KDN1}) will be evaluated using LES data in the next section.

Equation (\ref{eq:M_KDN1}) requires $\frac{\tau_{t0}}{\tau_{w0}}$, which is often unknown and unavailable as input for prediction of wind farm performance. \citet{kirby_ana2023} thus suggested to model this ratio as $\frac{\tau_{t0}}{\tau_{w0}} = 1 - H_F/h_0$, which stems from assuming that the (undisturbed) ABL streamwise shear stress is linear from the ground to the ABL height, $h_0$. With this model, one obtains a second simpler version
\begin{equation}
    M_{\rm KDN2} = \frac{1 + \frac{h_0}{L C_{f0}}  \left( 1 - \beta^2   \right)   }{\beta}, \label{eq:M_KDN2}
\end{equation}
which for example has been used by \citet{nishino_power_2025}.

Some further linearizations were given in the appendix of \citet{kirby_ana2023} to produce a third version
\begin{equation}
M_{\rm KDN3}  = 1 + \underbrace{\left(1.18  +   2.18 \frac{h_0}{L C_{f0}}    \right)}_{\zeta_{\rm KDN3}} (1 - \beta) ,  
\label{eq:M_KDN3}
\end{equation}
which is in a linear form similar to (\ref{eq:M_linear}). The model correctly captures that the momentum response factor, $\zeta_{\rm KDN3}$, should decrease with increasing farm size, as discussed earlier, through its dependence on the farm length $L$. Various quasi-1D assumptions were used to obtain (\ref{eq:M_KDN1}), which is why the farm size is represented by $L$ rather than the farm surface area $S_{\rm cv}$. 

\begin{figure}
  \centerline{\includegraphics[width=1\textwidth]{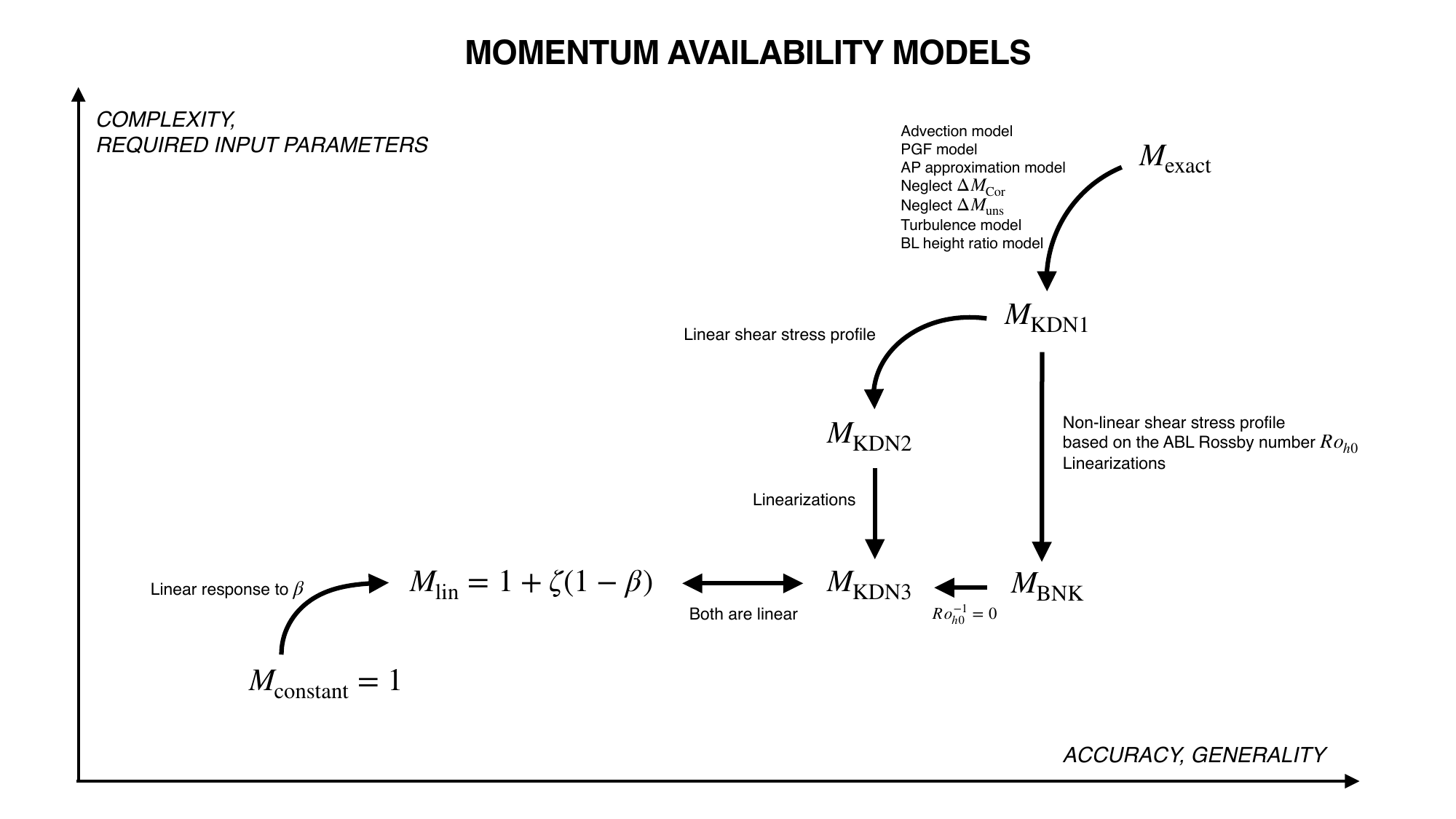}}
  \caption{Hierarchy of momentum availability models. We propose a new model, $M_{\rm BNK}$, in this paper, see § \ref{sec:rossby}. The arrows indicate assumptions to move from one model to another.}
\label{fig:hierarchy}
\end{figure}

An overview of the various momentum availability models is given in figure \ref{fig:hierarchy}, where they are, roughly, categorized according to their accuracy, generality, complexity and the amount of required input. In the next section, these models will be tested using LES data.

\section{Validation of momentum availability models with LES data}
\label{sec:test_of_M}

The momentum availability models presented in the previous section can be tested directly against the true value from LES data. In such a test, the ground truth, $M_{\rm exact}$, is obtained by evaluating (\ref{eq:M_contributions}) from the time-averaged 3D flow fields in the control volume surrounding the wind farm. Hence, full 3D fields of time-averaged velocity, pressure and Reynolds stresses are necessary for such a test.

\begin{figure}
  \centerline{\includegraphics[width=1.0\textwidth]{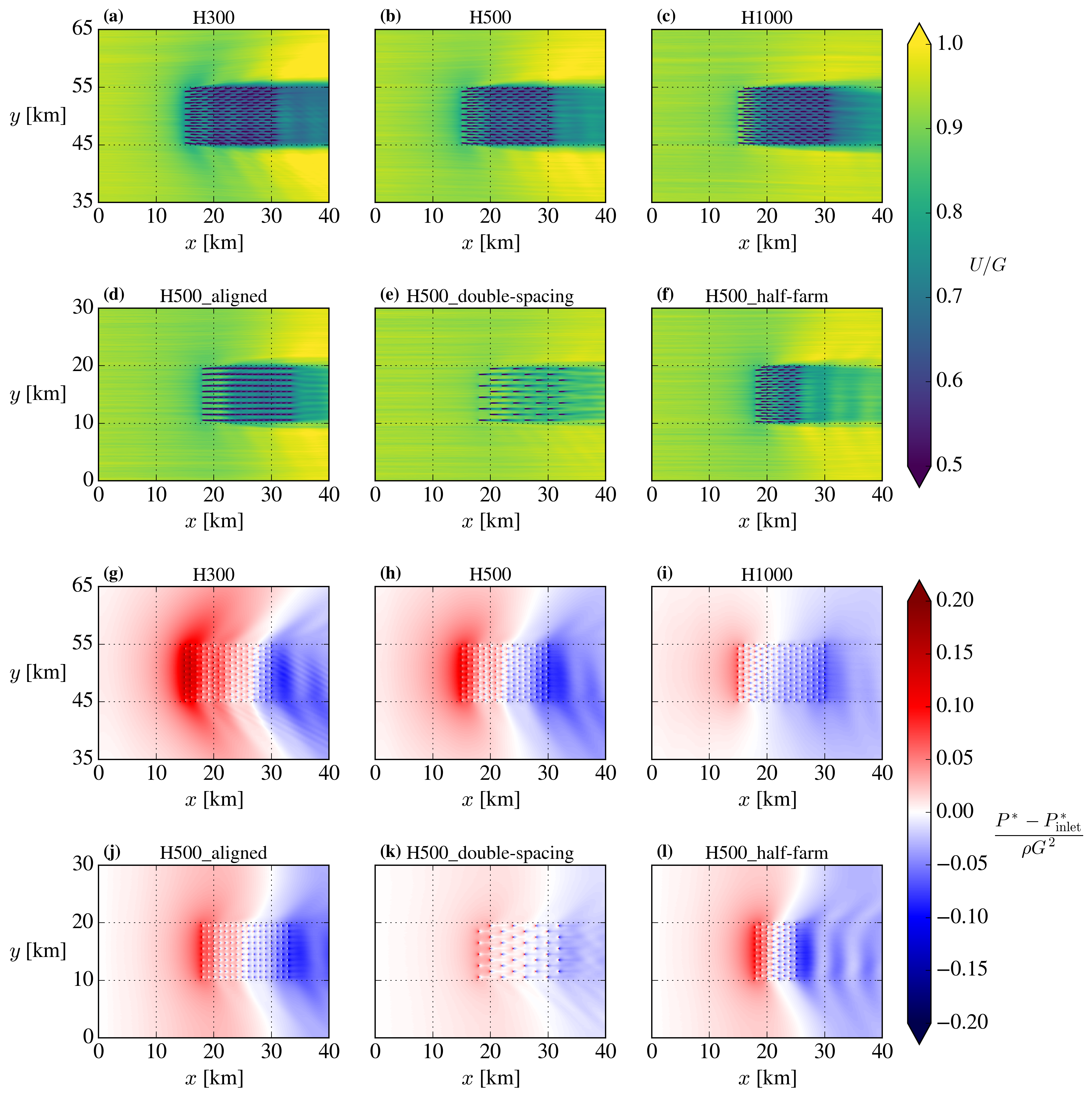}}
  \caption{Contours at hub-height of the time-averaged \textbf{(a-f)} streamwise velocity and \textbf{(g-l)} pressure perturbation (subtracted by inlet value). All cases have $N_t = 160$ turbines, except the double-spacing and half-farm cases, which have $N_t = 40$ and $N_t = 80$, respectively. The inter-spacing is $s_x/D = s_y/D = 5$, except in the double-spacing case. A staggered arrangement is used for all cases, except the aligned case. The farm width ($W \equiv N_y s_y$) is the same for all cases, while the farm length ($L \equiv N_x s_x$) is reduced for the half-farm case. This results in an array density of $\lambda = N_t A_d/S_{\rm cv} = 0.0314$ for all cases, except the double-spacing case, which has $\lambda = 0.00785$.}
\label{fig:hub_height_contours}
\end{figure}

In this study, we consider six cases from the LES database by \citet{lanzilao2024,lanzilao2025}. Three cases with the same farm layout but varying ABL heights, and three cases with the same ABL height but different wind farm layouts, are selected. A precursor-successor methodology was used by \citet{lanzilao2024,lanzilao2025} to simulate the wind farms in conventionally neutral boundary layers (CNBLs). The considered CNBLs are initialized with capping inversion strength $\Delta \theta = 5~{\rm K}$ and free lapse-rate $\Gamma = 4~{\rm K/km}$ (denoted with suffix `C5-G4' in the database), and three different initial boundary-layer heights $H=\{300,500,1000\}~{\rm m}$. All cases have geostrophic velocity $G = 10$ m/s, Coriolis frequency $f_c = 1.14 \times 10^{-4}$ s$^{-1}$ and aerodynamic roughness length $z_0 = 1 \times 10^{-4}$ m. Each wind turbine was modelled as an actuator disk (AD) with rotor diameter $D = 198$ m, hub-height $z_h = 119$ m, and a uniformly distributed thrust with thrust-coefficient $C_T'=1.94$ (based on the disk-averaged velocity). After the main simulation had spun up and the flow had reached a quasi-stationary state, the flow statistics were gathered over a time period to obtain first- and second-order statistics of the flow variables throughout the domain. As an example, the time-averaged flow fields at hub-height for the six cases are shown in figure \ref{fig:hub_height_contours}, from which it is clear that both the ABL height and layout have a large influence on the wind farm flow. For more details about the LES setup, we refer to \citet{lanzilao2024,lanzilao2025}.

\begin{figure}
  \centerline{\includegraphics[width=1.0\textwidth]{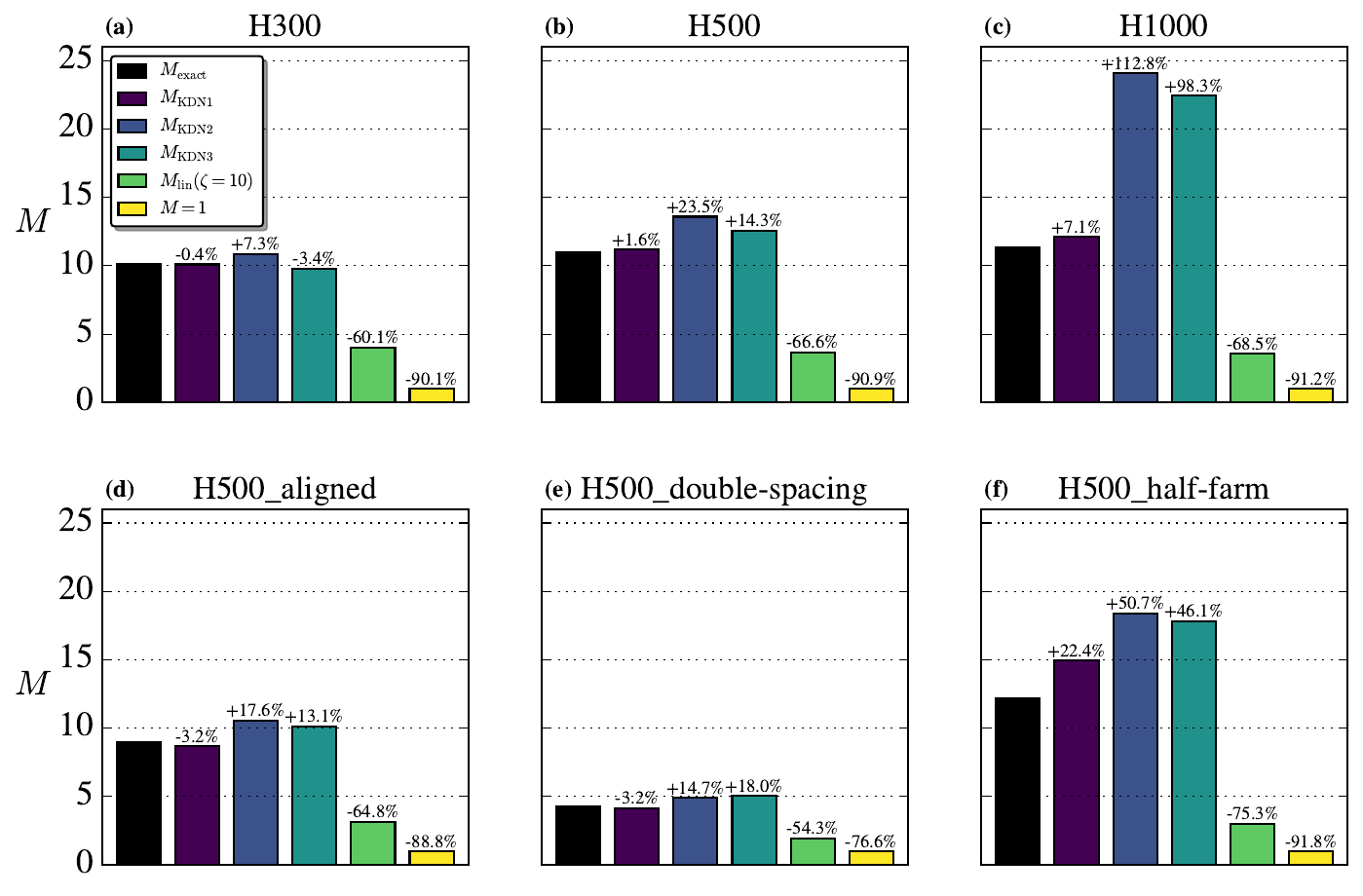}}
  \caption{Test of existing momentum availability models with LES data.}
\label{fig:test_of_M}
\end{figure}

Figure \ref{fig:test_of_M} shows the test of the various momentum availability models described in the previous section. The highest fidelity model $M_{\rm KDN1}$ is within $\pm 10\%$ of $M_{\rm exact}$ for five of the six cases, but also requires the most input parameters. The simpler $M_{\rm KDN2}$ and $M_{\rm KDN3}$ models perform moderately with predictions ranging from $-5\%$ to $25\%$ relative to $M_{\rm exact}$ for four of the six cases, but with a large overprediction for the half-farm and, most severely, for the H1000 case. In general, the $M_{\rm KDN}$ models appear to decrease in accuracy for increasing ABL heights. For completeness, the empirical linear model with $\zeta = 10$ (the typical order of magnitude, c.f. \citet{Patel2021}) and the no-response model ($M = 1$) are also included, but are clearly not performing well and should be avoided. In the next section, we will quantify the contributions of the various momentum mechanisms to $M$ and test the sub-models (\ref{eq:Madv_model}-\ref{eq:Muns_model}) that constitute the $M_{\rm KDN}$ models. 

\section{Test of sub-models in the $M_{\rm KDN}$ models}
\label{sec:LES}
To understand the performance of the $M_{\rm KDN}$ models, we will directly evaluate each constituent part of $M$ in this section, i.e. each term in (\ref{eq:M_contributions}), from the LES data. It was identified in the previous section that the accuracy of all three versions of the $M_{\rm KDN}$ model decreases for increasing ABL height, and we will thus limit the investigations in this section to the H300, H500 and H1000 cases.

\subsection{General observations relevant for the two-scale momentum theory}
\label{subsec:general_observations}

\begin{table}
  \begin{center}
\def~{\hphantom{0}}
  \begin{tabular}{lccccccccc}
              & $h_0$ [m]   & $p$ & $U_{F0}/G$ & $C_{f0}$ &  $\tau_{t0}/\tau_{w0}$ \\[3pt]
H300 & $357$ & $1.20$ & $0.96$ & $0.00177$ & $0.091$\\
H500 & $552$ & $1.28$ & $0.95$ & $0.00181$ & $0.329$\\
H1000 & $1095$ & $1.93$ & $0.93$ & $0.00183$ & $0.427$\\
  \end{tabular}
  \caption{Integral quantities for the LES precursor simulations. The quantities are the ABL height $h_0$, concavity of the total shear stress profile $p$, normalized mean streamwise velocity in the CV without turbines $U_{F0}/G$, friction coefficient $C_{f0}$ and the ratio of streamwise shear stress at the top of the CV to streamwise wall shear stress $\tau_{t0}/\tau_{w0}$.
  }
  \label{tab:les_precursor_integral_quantities}
  \end{center}
\end{table}
\begin{table}
  \begin{center}
\def~{\hphantom{0}}
  \begin{tabular}{lcccccccccc}
              & $\beta$  & $\beta_{\rm local}(0)$ & $\beta_{\rm local}(L)$   & $\gamma$ & $C_T^*$ & $\Delta P^*/(\rho G^2)$ \\[3pt]
H300 & $0.70$ & $0.85$ & $0.68$ & $1.28$ & $1.09$ & $-0.184$\\
H500 & $0.73$ & $0.88$ & $0.72$ & $1.13$ & $1.08$ & $-0.177$\\
H1000 & $0.74$ & $0.94$ & $0.69$ & $1.02$ & $1.08$ & $-0.115$\\
  \end{tabular}
  \caption{Integral quantities for the LES wind farm simulations. The quantities are the wind-speed reduction factor $\beta$, the local wind-speed reduction factor at the start of the farm $\beta_{\rm local}(0)$, the local wind-speed reduction factor at the end of the farm $\beta_{\rm local}(L)$, the friction exponent $\gamma$, the internal thrust coefficient $C_T^*$ and the normalized perturbation pressure jump over the farm $\Delta P^*/(\rho G^2)$.
  }
  \label{tab:les_main_integral_quantities}
  \end{center}
\end{table}

Some integral quantities of the LESs relevant for the two-scale momentum theory are shown in table \ref{tab:les_precursor_integral_quantities}-\ref{tab:les_main_integral_quantities}. The boundary-layer height $h_0$ and concavity $p$ have been estimated through a fit of the normalized total shear stress to $|\tau|_0/|\tau|_{w0} = (1 - z/h_0)^p$. A discussion about the shape of the ABL shear stress profiles and its relevance for the two-scale momentum theory will be given in § \ref{sec:rossby}. The values of the friction coefficient $C_{f0}$ are within the typical expected range for offshore wind farms.

\begin{figure}
    \hspace{-20pt}
  \centerline{\includegraphics[width=0.9\textwidth]{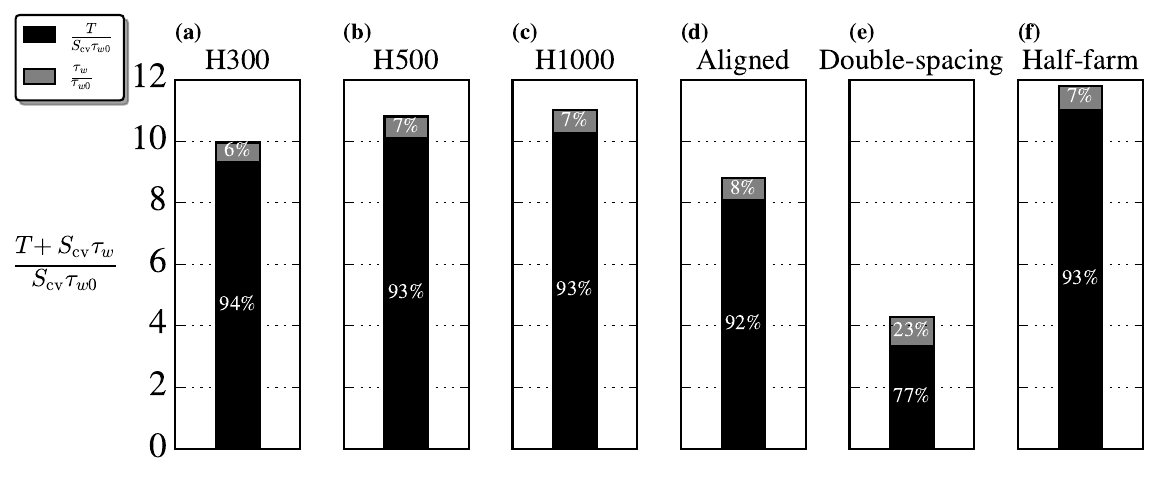}}
  \caption{Momentum sinks in the NDFM equation, i.e. left-hand side of (\ref{eq:NDFM}) and (\ref{eq:NDFM_LHS}). For completeness, the momentum sinks in the aligned, double-spacing and half-farm cases are also included, although these cases will not be further investigated in this section.}
\label{fig:momentum_sinks}
\end{figure}

In earlier studies, it has generally been assumed that the surface friction exponent, $\gamma$, is around 1.7--1.8 \citep[e.g.][]{Kirby2022}, but the current LESs, see table \ref{tab:les_main_integral_quantities}, show that it can be significantly smaller, i.e. around 1.0--1.3. A smaller value of $\gamma$ means that the wall friction contribution, i.e. $\beta^\gamma = \tau_w/\tau_{w0}$, in the NDFM equation (\ref{eq:NDFM}), becomes larger (because $0 < \beta < 1$). Nevertheless, as shown in figure \ref{fig:momentum_sinks}, the farm thrust term, i.e. $C_{T}^* \frac{\lambda}{C_{f0}} \beta^2 = T/(S_{\rm cv} \tau_{w0})$, is still the dominant momentum sink on the left-hand side of NDFM equation for all cases. This is an important observation, because it implies that the modelling of $\gamma$ is not critical to the prediction of $\beta$. One can therefore, in many cases, assume $\gamma = 2$ to simplify the NDFM equation without incurring any significant error, even if the true value of $\gamma$  is closer to one. On the other hand, the modelling of the internal thrust coefficient, $C_T^*$, is more important for obtaining good predictions of $\beta$ with the NDFM equation. Regarding $C_T^*$, table \ref{tab:les_main_integral_quantities} shows that it is nearly constant for all three simulations, which is consistent with the fundamental two-scale assumption in the sense that it is mainly an `internal' parameter of the wind farm and therefore is not expected to be sensitive to the ABL height, which is an `external' parameter \citep{kirby_turbine-_2025}.

\begin{figure}
  \centerline{\includegraphics[width=1\textwidth]{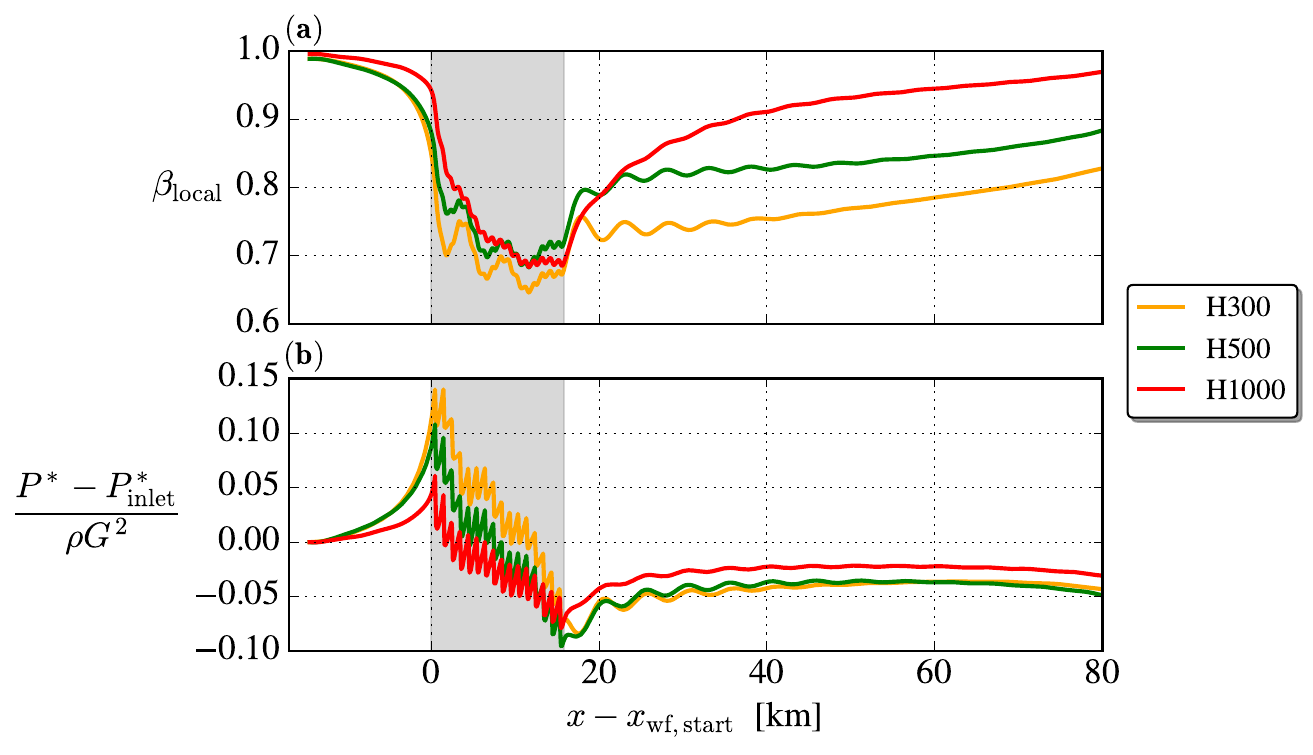}}
  \caption{\textbf{(a)} Local wind-speed reduction factor and \textbf{(b)} normalized perturbation pressure (subtracted by inlet value) spanwise-averaged over the width $y=[y_{\rm center}-W/2,y_{\rm center}+W/2]$ and height $z=[0,H_F]$ of the CV. The CV-region enclosing the wind-farm is marked with gray.}
\label{fig:beta_pressure_development}
\end{figure}

The development of the local wind-speed reduction factor $\beta_{\rm local}$ and the perturbation pressure $P^* \equiv P - P_\infty$ (where $P_\infty$ is the background pressure, which is linearly decreasing in the horizontal direction according to the driving PGF, $\partial P_\infty/\partial x_i$, of the simulation) are shown in figure \ref{fig:beta_pressure_development}. The small fluctuations in the farm-region are from the individual turbines, whereas the larger fluctuations, which are also present in the farm-wake, are from farm-induced gravity waves. Gravity waves are also clearly visible in figure \ref{fig:hub_height_contours} and have been studied in detail by \citet{lanzilao2024} among others. They are generally stronger for lower ABL heights and it is also seen in figure \ref{fig:beta_pressure_development}(b) that a lower ABL height induces a larger pressure perturbation drop, $\Delta P^*$, over the farm region, which is favorable for farm power production. The pressure perturbation, $P^*$, should go to the same constant value, $P_{\rm inlet}^*$, far upstream and downstream of the farm, but it does not despite the fact that a large domain was used for the LESs, which suggests that an even longer domain should ideally have been used. An error can be expected from this in the PGF term of the momentum availability factor, but this is not expected to change the main conclusions of the following analysis.

\subsection{Advection}

\begin{figure}
  \centerline{\includegraphics[width=1\textwidth]{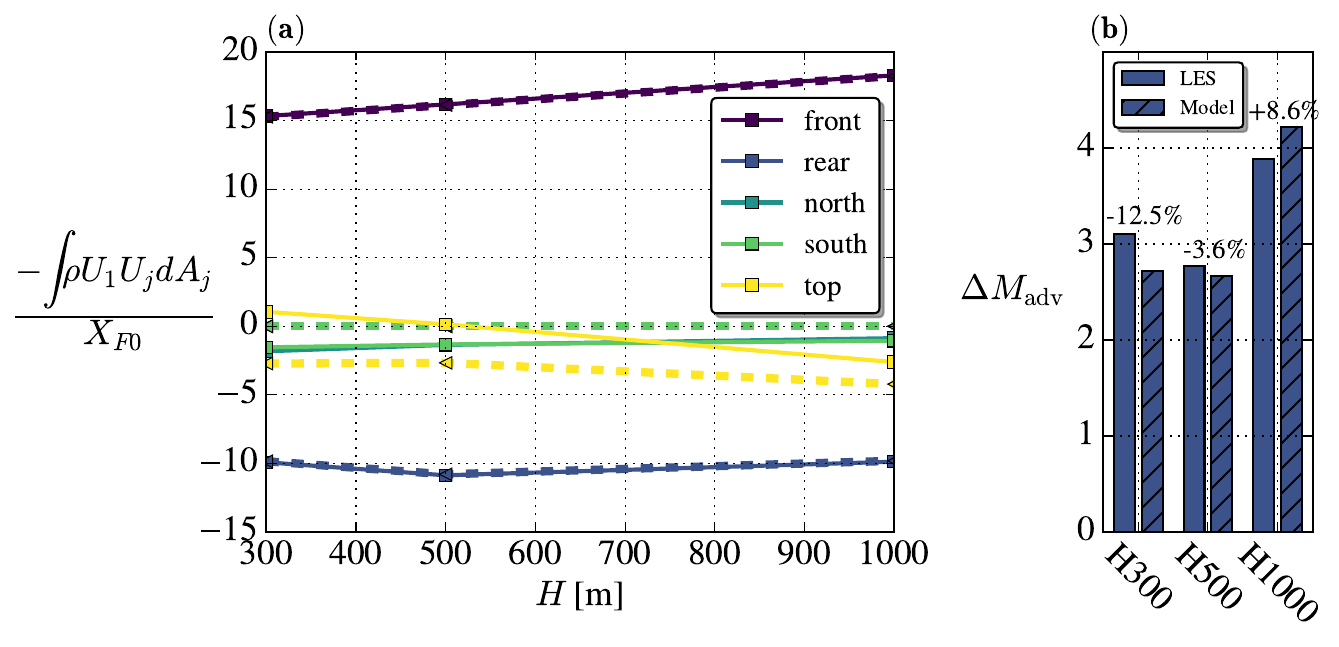}}
  \caption{\textbf{(a)} Contributions from CV surfaces to the advection term for  LES (full) and  model (dashed). Positive sign means advection of $U_1$ into the CV. Note that the north and south contributions are almost identical. \textbf{(b)} Total advection contribution and error of the model.}
\label{fig:advection_parts}
\end{figure}

Streamwise momentum is advected in and out of the CV, and its contribution, $\Delta M_{\rm{adv}}$, to the momentum availability factor is defined in (\ref{eq:M_contributions}). For our current cases, the expression for the advection part simplifies to
\begin{align}
\Delta M_{\rm{adv}} &=  \frac{X_{\rm adv}}{X_{F0}} = \frac{-\int_{\Omega_{\rm cv}}  \rho U_1 U_j dA_j}{X_{F0}} ,
\end{align}
because $X_{\rm adv0} = 0$ due to the horizontal homogeneity of the precursor simulations (the front/rear and north/south advection cancel due to periodicity and the top/bottom advection are zero because $U_3 = 0$). There is also no flow through the bottom surface in the main simulation, hence only the front/rear, north/south, and top surfaces of the CV contribute to the integral, which are calculated directly from the LES flow field data and displayed in figure \ref{fig:advection_parts}(a). In the advection model of \citet{kirby_ana2023}, i.e. (\ref{eq:Madv_model}), it was assumed that the north/south contributions are zero, but as shown in figure \ref{fig:advection_parts}(a), both of these are in fact slightly negative for the current cases (meaning there is advection out of the CV at these surfaces) due to the flow through the farm diverting horizontally spanwise. However, in comparison with the other contributions, the north/south contributions are indeed small and it is therefore a reasonable approximation to neglect them. The front/rear contributions used in their model are exact, as they can be rewritten as
\begin{equation}
    - \frac{1}{X_{F0}} \int_{\Omega_{\rm front}}  \rho U_1 U_j dA_j = \frac{2 \beta_{\rm local}(0)^2 H_F}{L C_{f0}} \label{eq:deltaMadv_frontmodel} ,
\end{equation}
\begin{equation}
    - \frac{1}{X_{F0}} \int_{\Omega_{\rm rear}}  \rho U_1 U_j dA_j = - \frac{2 \beta_{\rm local}(L)^2 H_F }{L C_{f0}}  \label{eq:deltaMadv_rearmodel}.
\end{equation}
The sign change on the former is because the vectors $U_j$ and $dA_j$ have opposite directions on the front face. Finally, for the top surface \citet{kirby_ana2023} used a simple quasi-1D mass flux analysis to estimate its contribution as
\begin{align}
    - \frac{1}{X_{F0}} \int_{\Omega_{\rm top}}  \rho U_1 U_j dA_j &= \frac{H_F \left( \beta_{\rm local}(L)^2 - \beta_{\rm local}(0)^2 \right)}{L C_{f0}}  . \label{eq:deltaMadv_topmodel}  
\end{align}
In all cases, $\beta_{\rm local}(L) < \beta_{\rm local}(0)$, hence the model always predicts advection out of the top surface (i.e. a negative contribution). The LES data in figure \ref{fig:advection_parts}(a) however show that there is advection into the CV (i.e. a positive contribution) for the lowest ABL height case, H300, which can occur due to the strong gravity waves present in this case. The model for the top surface advection (\ref{eq:deltaMadv_topmodel}) overpredicts the advection out of the top surface in all cases, which in effect compensates for the neglected advection out of the north/south surfaces. The overall advection model (\ref{eq:Madv_model}) is the sum of (\ref{eq:deltaMadv_frontmodel}-\ref{eq:deltaMadv_topmodel}), and figure \ref{fig:advection_parts}(b) shows it is within approximately $\pm 10\%$ of the LES data.

\subsection{Pressure gradient forcing}

The CNBLs used in the current cases are driven by a horizontal pressure gradient forcing, $(\partial P_\infty/\partial x_1,\partial P_\infty/\partial x_2)$, which is constant throughout the domain, i.e. it is a barotropic ABL, and, as shown in figure \ref{fig:beta_pressure_development}(b), a perturbation PGF, $\partial P^*/\partial x_1$, develops in the CV region due to the presence of the wind farm (recall that $P^* \equiv P - P_\infty$ is defined as the pressure perturbation from the linearly decreasing background pressure $P_\infty$ that drives the flow). The contribution from the pressure gradient to the momentum availability is defined in (\ref{eq:M_contributions}) and the expression for the PGF part can be rewritten with the perturbation pressure as
\begin{align}
    \Delta M_{\rm{PGF}} &= -\frac{V_{\rm cv}}{X_{F0}}  \left[  \frac{\partial P^*}{\partial x_1}\right] = - \frac{H_F}{L \tau_{w0}} \underbrace{\left( P_{\rm rear}^* - P_{\rm front}^* \right)}_{\equiv \Delta P^*} ,\label{eq:Mpgf_front_rear}
\end{align}
where $P_{\rm rear}^*$ and $P_{\rm front}^*$ are the rear and front surface-averaged values, respectively. The above expression is exact and was the starting point for \citet{kirby_ana2023}, who used Bernoulli's equation to estimate the front perturbation pressure as
\begin{equation}
    P_{\rm front}^* - P_{\rm inlet}^* = \frac{1}{2} \rho U_{F0}^2 \left( 1 - \beta_{\rm local}(0)^2\right) ,
    \label{eq:Mpgf_front_model}
\end{equation}
where $P_{\rm inlet}^*$ is the far upstream value. The second assumption of their PGF model is that the rear perturbation pressure is simply the negative value of the above, i.e.
\begin{equation}
    P_{\rm rear}^* - P_{\rm inlet}^* = -(P_{\rm front}^* - P_{\rm inlet}^*) ,
    \label{eq:Mpgf_rear_model}
\end{equation}
which corresponds to assuming that $P^* - P_{\rm inlet}^*$ is anti-symmetric around the center of the farm. Combining (\ref{eq:Mpgf_rear_model}) with (\ref{eq:Mpgf_front_rear}-\ref{eq:Mpgf_front_model}) gives the final model of \citet{kirby_ana2023}, equation (\ref{eq:Mpgf_model}).

\begin{figure}
  \centerline{\includegraphics[width=1\textwidth]{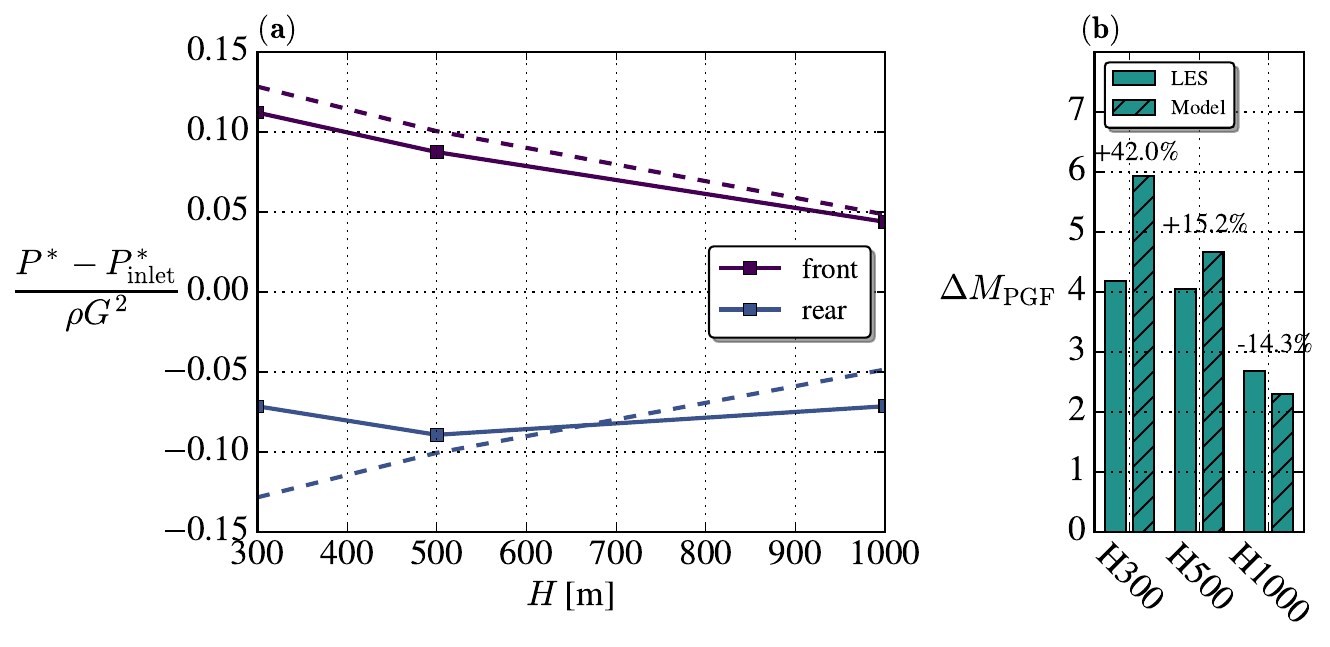}}
  \caption{\textbf{(a)} Contributions from the front and rear surfaces to the PGF term for LES (full) and  model (dashed). \textbf{(b)} Total PGF contribution and error of the model.}
\label{fig:pgf_assumptions}
\end{figure}

The two assumptions, i.e. (\ref{eq:Mpgf_front_model}) and (\ref{eq:Mpgf_rear_model}), are checked with the LES data in figure \ref{fig:pgf_assumptions}(a), which shows that the simple Bernoulli estimate is fairly well capturing the trend of the front pressure, whereas the rear assumption is less consistent. Regarding the latter, it can indeed be observed in figure \ref{fig:beta_pressure_development}(b) that the perturbation pressure is not perfectly anti-symmetric around the center of the farm. The H500 case is closest to being anti-symmetric, which is consistent with that this is the case with the best rear prediction. As discussed at the end of § \ref{subsec:general_observations}, the LES domain length is possibly too short to simulate the pressure field around the farm accurately, although it is not clear how large the impact of the LES domain length is and if a longer inlet fetch would improve the agreement between the LES and model. 

The difference between the two dashed lines (or full lines) in figure \ref{fig:pgf_assumptions}(a) is the pressure perturbation jump, $\Delta P^*$, over the farm, which per (\ref{eq:Mpgf_front_rear}) determines the PGF contribution to $M$. It can be seen that the pressure jump is overestimated by the model for the H300 and H500 cases, which is why $\Delta M_{\rm{PGF}}$ is overestimated for these two cases, see figure \ref{fig:pgf_assumptions}(b). Nevertheless, the model correctly captures that the PGF contribution should decrease with increasing ABL height, which is expected because the ABL top acts as a `lid' on the flow and thereby affects the pressure gradient within the farm (a taller ABL height gives a smaller enhancement of the favorable wind farm pressure gradient). The effect of the ABL height on the wind farm pressure field has also been studied in detail by \citet{lanzilao2024}.

\subsection{The AP approximation}
Combining the advection and PGF models (\ref{eq:Madv_model}-\ref{eq:Mpgf_model}) gives
\begin{align}
    \Delta M_{\rm{adv}} + \Delta M_{\rm{PGF}} 
    &= \frac{H_F}{L C_{f0}}  \left( - \beta_{\rm local}(0)^2 - \beta_{\rm local}(L)^2 + 2   \right) ,
    \label{eq:deltaMadvpgf_sum}
\end{align}
which can be simplified to (\ref{eq:Madv_and_Mpgf_sa}) by invoking the `AP approximation'
\begin{equation}
    \beta_{\rm local}(0)^2 + \beta_{\rm local}(L)^2 \approx 1 + \beta^2     . \label{eq:sa}
\end{equation}
This approximation was proposed by \citet{kirby_ana2023} by assuming that the decrease of $\beta_{\rm local}$ from upstream (where it has a value of 1) to the front of the farm is equal to the increase from the farm-averaged value (i.e. $\beta$) to $\beta_{\rm local}$ at the end of the farm, i.e.
\begin{equation}
    1 - \beta_{\rm local}(0) \approx \beta_{\rm local}(L) - \beta ,  \quad \textrm{assumption (i)},\label{eq:sa_assumption1}
\end{equation}
which when squared and rearranged becomes
\begin{align}
    1 + \beta^2    &\approx \beta_{\rm local}(L)^2 + \beta_{\rm local}(0)^2 + 2 \left( \beta_{\rm local}(L) \beta_{\rm local}(0) - \beta \right) . \label{eq:sa_squared}
\end{align}
This equation simplifies to (\ref{eq:sa}), when the second assumption is invoked
\begin{align}
     2 \left( \beta_{\rm local}(L) \beta_{\rm local}(0) - \beta \right) \approx 0,  \quad \textrm{assumption (ii)} . \label{eq:sa_assumption2}
\end{align}

\begin{figure}
  \centerline{\includegraphics[width=0.95\textwidth]{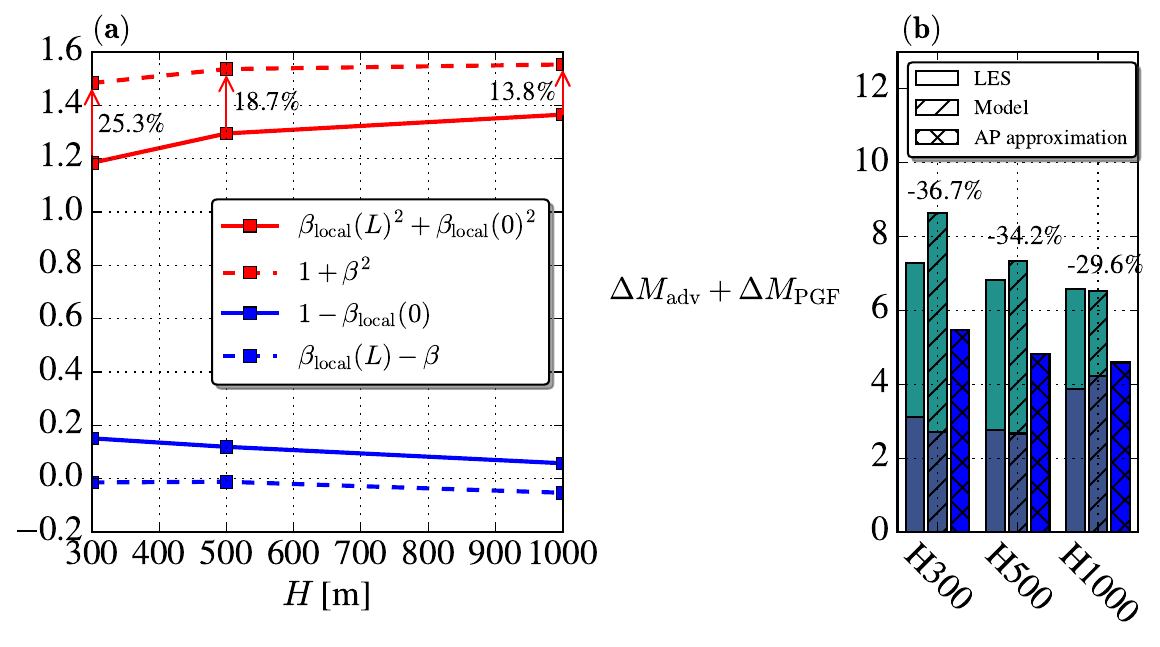}}
  \caption{\textbf{(a)} Test of the AP approximation (the red lines) and the underlying assumption (i) (\ref{eq:sa_assumption1}) (blue lines). \textbf{(b)} Combination of the advection and PGF terms of the momentum availability factor. The error percentage indicated is between model and AP approximation bars.}
\label{fig:sa_assumptions}
\end{figure}

The data from the LESs in figure \ref{fig:sa_assumptions}(a) show an absolute error of around 0.1--0.2 for assumption (i) (\ref{eq:sa_assumption1}). The error from this first assumption means that (\ref{eq:sa_squared}) is not exactly correct, which together with the error from the second assumption leads to that the AP approximation gives an overestimation of $\beta_{\rm local}(L)^2 + \beta_{\rm local}(0)^2$ by around 14--25\%. In contrast, \citet{kirby_ana2023} only found an overprediction of around 10\% for the four LES cases they analyzed. The overprediction of the AP approximation means that the sum of the advection and PGF terms gets underpredicted by 30--35\%, see figure \ref{fig:sa_assumptions}(b), because $\beta_{\rm local}(0)^2 + \beta_{\rm local}(L)^2$ appears with a negative sign in (\ref{eq:deltaMadvpgf_sum}). These results highlight the difficulty in accurately predicting the advection and PGF terms without using the local flow information (such as $\beta_{\rm local}$) as input.

\subsection{Coriolis}
\label{sec:coriolis}
It was assumed by \citet{kirby_ana2023} that the contribution from the Coriolis force to the momentum availability factor could be neglected, i.e. (\ref{eq:Mcor_model}). To verify the correctness of this, we evaluate its contribution directly from the LES data using its definition (\ref{eq:M_contributions}), which is
\begin{align}
    \Delta M_{\rm{Cor}} &= \frac{V_{\rm cv} \rho f_c}{X_{F0}}  \left( \left[U_2\right] - \left[U_2\right]_0\right) = \frac{2 }{C_{f0} Ro_{F0}} \left( \frac{\left[U_2\right] - \left[U_2\right]_0 }{U_{F0}}\right) ,
\end{align}
where $Ro_{F0} \equiv U_{F0}/(f_c H_F)$ is a Rossby number based on the CV height and no-turbine farm velocity.
All three simulations have $\frac{2 }{C_{f0} Ro_{F0}} \approx 4$ (which can be evaluated from the values in table \ref{tab:les_precursor_integral_quantities}), but the change of CV-averaged spanwise velocity from the precursor to main simulation is small, $\left( \frac{\left[U_2\right] - \left[U_2\right]_0 }{U_{F0}}\right) = \mathcal{O}(0.01)$, and therefore $\Delta M_{\rm{Cor}}$ is only around $0.05$, see table \ref{tab:deltaM_cor}. This is much smaller than the contributions from the advection and PGF terms (see figure \ref{fig:sa_assumptions}(b)); hence it is a good assumption to neglect the Coriolis contribution.

\begin{table}
  \begin{center}
\def~{\hphantom{0}}
  \begin{tabular}{l|c|c|c|c|}
            & $\frac{2 }{C_{f0} Ro_{F0}}$ & $\frac{\left[U_2\right] - \left[U_2\right]_0}{U_{F0}}$ & $\Delta M_{\rm Cor}$ (LES) & $\Delta M_{\rm Cor}$ (model)  \\[3pt]
H300 & $4.00$ & $0.02$ & $0.07$ & $0.00$\\
H500 & $3.97$ & $0.01$ & $0.05$ & $0.00$\\
H1000 & $3.99$ & $0.01$ & $0.04$ & $0.00$\\
  \end{tabular}
  \caption{LES and model predictions of the Coriolis contribution.}
  \label{tab:deltaM_cor}
  \end{center}
\end{table}

\subsection{Unsteadiness}
\label{sec:unsteady}

The effect of (farm-scale) unsteadiness on the momentum availability was assumed to be negligible in the model of \citet{kirby_ana2023}, i.e. $\Delta M_{\rm uns} = 0$ (\ref{eq:Muns_model}). To validate this, the time history of the flow is necessary, which is not available in the LES database of \citet{lanzilao2024,lanzilao2025}. As an alternative, we will instead consider the time- and horizontally-averaged streamwise momentum budgets to implicitly give an assessment of the stationarity of the flow in the present LESs.

\begin{figure}
  \centerline{\includegraphics[width=1\textwidth]{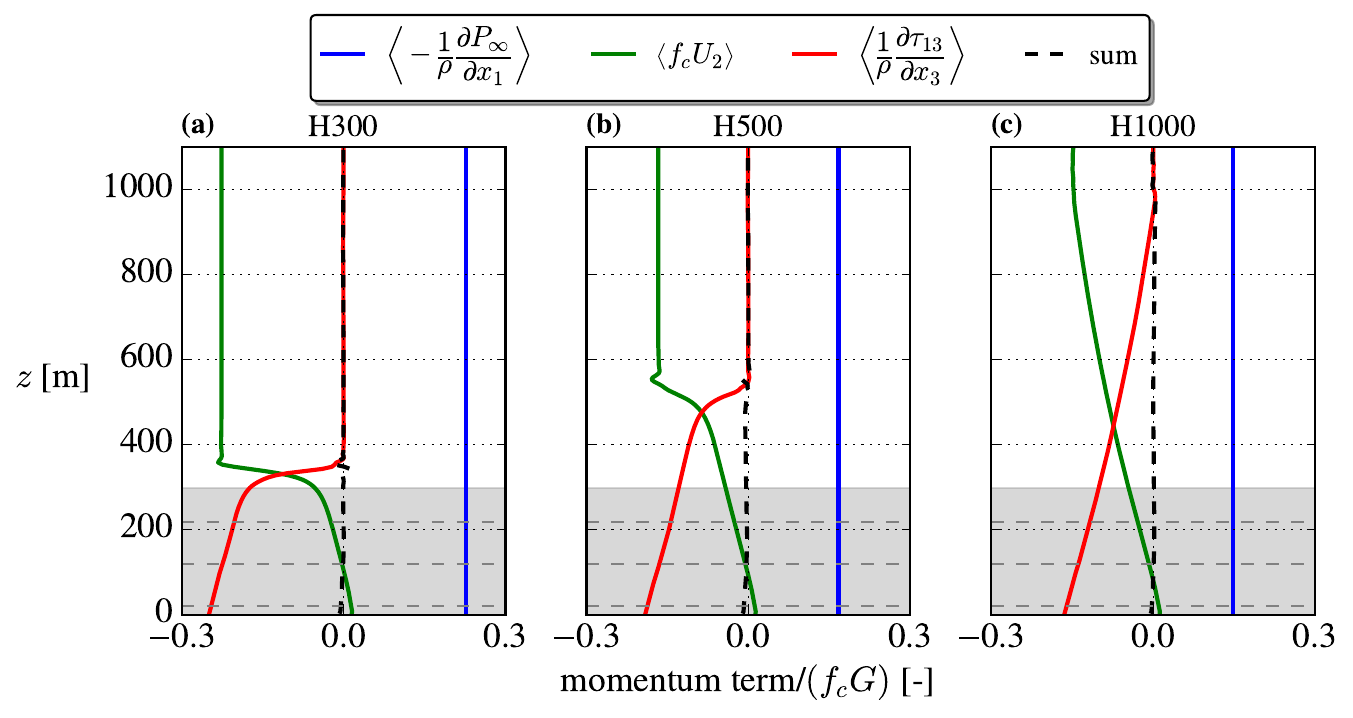}}
  \caption{Momentum budget for CNBL precursor simulations. Gray area marks CV region.}
\label{fig:momentum_balance_precursors}
\end{figure}

The momentum budget for a CNBL precursor simulation is given by
\begin{align}
      \left\langle\frac{\partial U_1}{\partial t}\right\rangle  =  \left\langle- \frac{1}{\rho} \frac{\partial P_\infty}{\partial x_1}\right\rangle  + \left\langle f_c  U_2\right\rangle + \left\langle\frac{1}{\rho} \frac{\partial {\tau_{13}}}{\partial x_3}\right\rangle,     \label{eq:mom_budget_precursor}
\end{align}
where $\langle \cdot \rangle$ is the horizontal-average operator. The advection term is zero and the stress divergence term is simplified due to the periodic BCs applied in the horizontal directions. A subscript $\infty$ is used on the pressure to denote that this is the driving pressure of the ABL. The budgets of the three considered ABLs are shown in figure \ref{fig:momentum_balance_precursors} together with the sum of the terms on the right-hand side of (\ref{eq:mom_budget_precursor}). At all heights, the sum is much smaller than the other terms, and we thus conclude that the ABLs are close to being stationary (i.e. `quasi-stationary'). Previous studies of the CNBL have indicated that around six non-dimensional time units are sufficient for the flow to reach a quasi-stationary state \citep{pedersen_structure_2014,LiuGDL2021,Liu2024} and \citet{lanzilao2025} indeed time-averaged the flow over the non-dimensional time period $f_c t = \left[10.5, 11.3\right]$.

\begin{figure}
  \centerline{\includegraphics[width=1\textwidth]{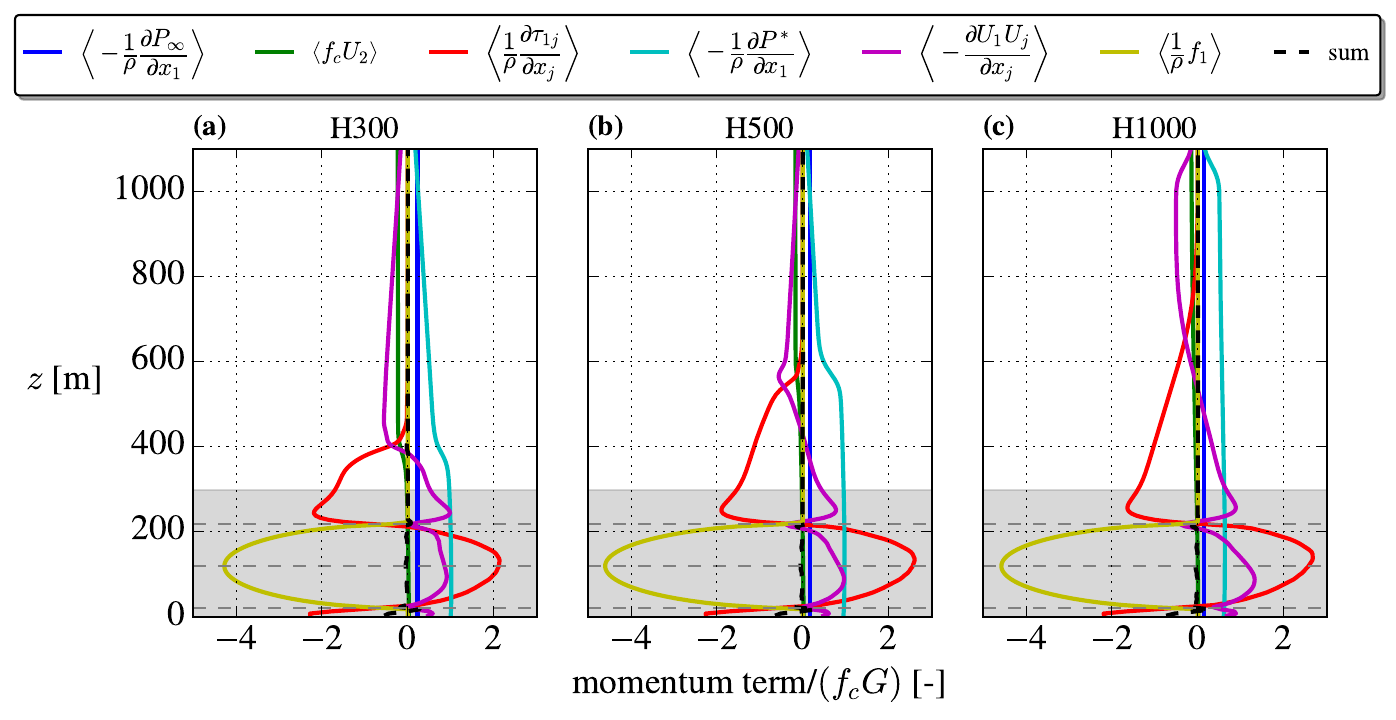}}
  \caption{Momentum budget for wind farm simulations. Gray area marks CV region.}
\label{fig:momentum_balance_cv}
\end{figure}

The momentum budget for a wind farm simulation is given by
\begin{align}
    \left\langle\frac{\partial U_1}{\partial t}\right\rangle =  \left\langle- \frac{\partial U_1 U_j }{\partial x_j}\right\rangle     + \left\langle -\frac{1}{\rho} \frac{\partial P^* }{\partial x_1}\right\rangle + \left\langle -\frac{1}{\rho} \frac{\partial P_\infty}{\partial x_1}\right\rangle + \left\langle f_c  U_2\right\rangle + \left\langle\frac{1}{\rho}\frac{\partial \tau_{1j}}{\partial x_j}\right\rangle + \left\langle\frac{1}{\rho} f_1\right\rangle , \label{eq:mom_budget}
\end{align}
where the horizontal-average operator $\langle \cdot \rangle$ is over the CV surface area, $S_{\rm cv}$, only. No body force fields are available in the analyzed LES dataset, but the body force term in (\ref{eq:mom_budget}) can nevertheless still be obtained, because the total farm thrust, $T$, is known, and because the thrust on each AD is distributed uniformly; hence, similar to the actuator wind farm (AWF) model \citep{vanderlaan2024awf}, we have that the force is distributed in the vertical direction as
\begin{equation}
 q_{\rm ver}(z) =  \begin{cases} 
         \frac{4}{\pi D} \sqrt{1 - ([z-z_H]/R)^2} &\quad \textrm{if} \quad |(z-z_H)/R| \le 1 \\
          0 &\quad \textrm{if} \quad |(z - z_H)/R| > 1 \\             
          \end{cases} ,
\end{equation}
and thus have
\begin{align}
    \left\langle\frac{1}{\rho}f_1\right\rangle  = q_{\rm ver}\frac{T}{S_{\rm cv} \rho}   .
\end{align}
Figure \ref{fig:momentum_balance_cv} shows the budgets for the three cases and again the sum of the terms is close to zero at all heights, thus signifying quasi-stationarity of the wind farm flows as well. In summary, we conclude that the transient evolution of the streamwise momentum is small in the present LESs and that the contribution of unsteadiness to the momentum availability can be neglected for all three cases.

\subsection{Boundary-layer height}
\label{sec:bl_height}
\begin{figure}
  \centerline{\includegraphics[width=1\textwidth]{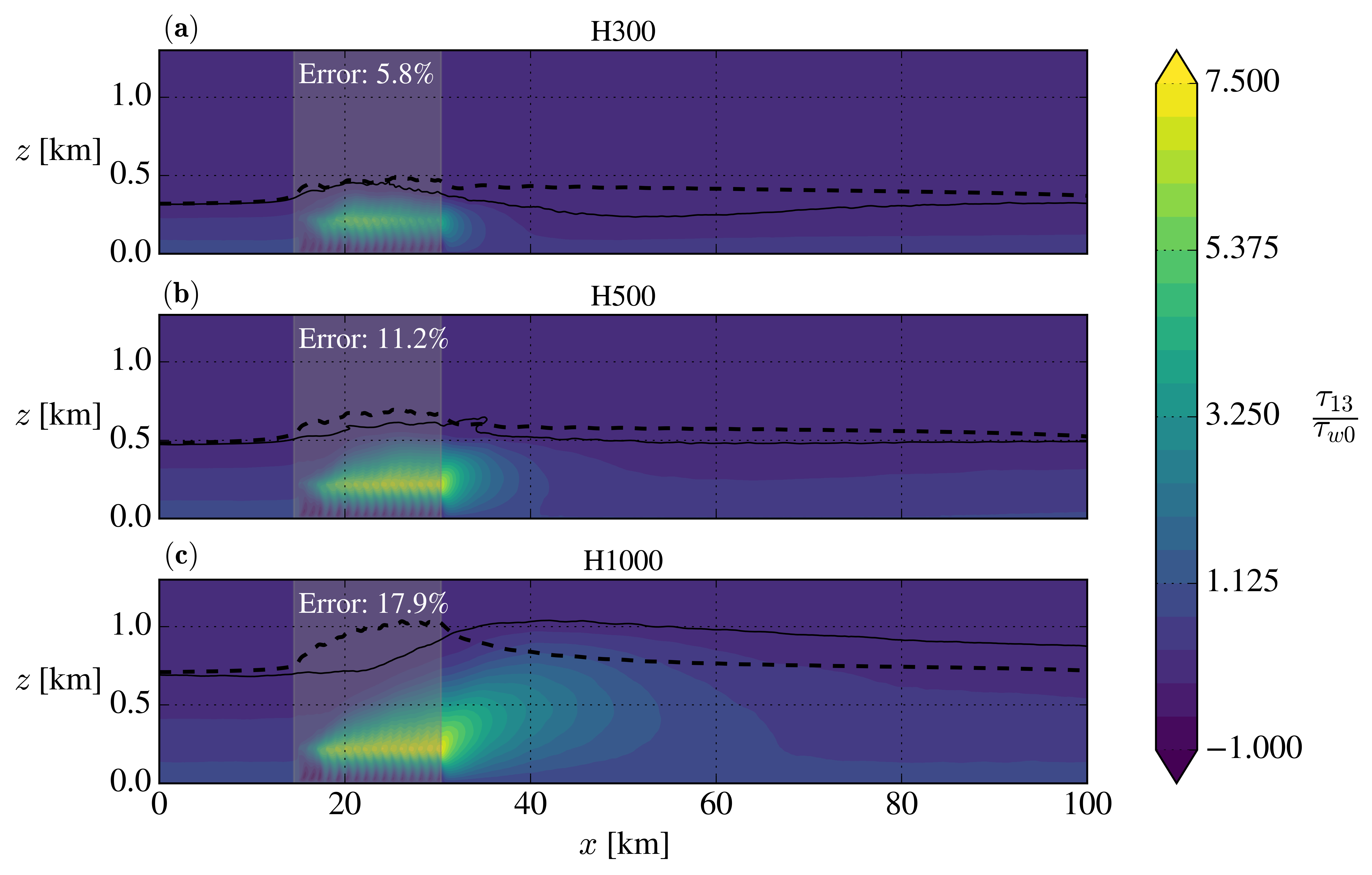}}
  \caption{Contours of the spanwise-averaged (over the CV width $y=[y_{\rm center}-W/2,y_{\rm center}+W/2]$) streamwise shear stress. BL height $\delta(x)$ from LES (\full) and model (\dashed). The farm region is highlighted with a grey box and the model error, $(h^{\rm model} - h^{\rm LES})/h^{\rm LES}$, is shown.}
\label{fig:tau13_contour}
\end{figure}

\citet{kirby_ana2023}'s model of the momentum availability contribution from turbulence, $\Delta M_{\rm turb}$ (\ref{eq:Mstr_model}), depends on the height of the undisturbed ABL $h_0 \equiv \delta(x=-\infty)$ and the farm BL $h \equiv L^{-1} \int_{x_{\rm wf,start}}^{x_{\rm wf,start}+L} \delta(x) dx$. There are several ways to define the boundary-layer height, $\delta(x)$, and in this section we will define it as the height at which the shear stress, $\tau_{13}$, is equal to 5\% of its ABL surface value $\tau_{w0}$ \citep[e.g.][]{Kosovic2000}. Figure \ref{fig:tau13_contour} shows the spanwise-averaged streamwise shear stress from the three cases and highlights the contour line corresponding to the BL height. A large shear stress region develops in and above the wind farm in all cases due to the large velocity gradient, $\partial U / \partial z$, induced by the turbines, and this region eventually reaches the capping inversion layer and pushes it upwards until the buoyant destruction from the stable stratification of the free atmosphere is sufficiently large to suppress the turbulence. It is this complex balance that dictates the spatial evolution of $\delta(x)$.

\citet{kirby_ana2023} proposed a simple model for the spatial development of the boundary-layer height from a quasi-1D continuity analysis
\begin{equation}
    \delta^{\rm model}(x) = \frac{h_0}{\beta_{\rm local}(x)} ,
    \label{eq:bl_model}
\end{equation}
which is shown with dashed lines in figure \ref{fig:tau13_contour}. It can be seen that the model correctly captures an increase of the BL height in the farm region, but that it is less accurate for taller ABLs. In the H1000 case, the BL height first starts to increase after approximately a half farm length, which is in line with the fact that the shear stress region develops at an angle (not directly upwards from the wind farm entrance) and that for taller ABLs it thus takes a longer streamwise distance to reach the ABL height. The inability of the model to capture this is due to its quasi-1D assumption and that the model is derived from mass continuity, which implicitly implies that the BL height is directly related to the velocity field, i.e. that $\delta(x)$ follows the `separating' streamline between the ABL and the free atmosphere, rather than the shear stress field.

To obtain a model for the farm BL height, $h$, \citet{kirby_ana2023} used $\delta^{\rm model}(x)$ (\ref{eq:bl_model}), in the definition of $h$ described in the beginning of the current subsection
\begin{equation}
    h^{\rm model} = \frac{1}{L} \int_{x_{\rm wf,start}}^{x_{\rm wf,start}+L} \delta^{\rm model}(x) dx = \frac{h_0}{L} \int_{x_{\rm wf,start}}^{x_{\rm wf,start}+L} \frac{1}{\beta_{\rm local}} dx \approx \frac{h_0}{\beta} ,
\end{equation}
and assumed that the last integral is equal to $L/\beta$, which gives the final simple model, i.e. (\ref{eq:bl_height_ratio_model}). The latter `integral assumption' was confirmed by \citet{kirby_ana2023} with four LES cases from \citet{wu2017} to be very good (within 0.5\% of the actual value) and we find the same for the cases studied in this paper. This means that the error of $h^{\rm model}$ compared to $h^{\rm LES}$, which is on the order of 6--18\% (see figure \ref{fig:tau13_contour}), is almost entirely from (\ref{eq:bl_model}). This overprediction of $h$ has a significant influence on the model prediction of $\Delta M_{\rm turb}$, which we will return to in the next section.

\subsection{Turbulence}
\label{sec:stress_model}

\begin{figure}
  \centerline{\includegraphics[width=1\textwidth]{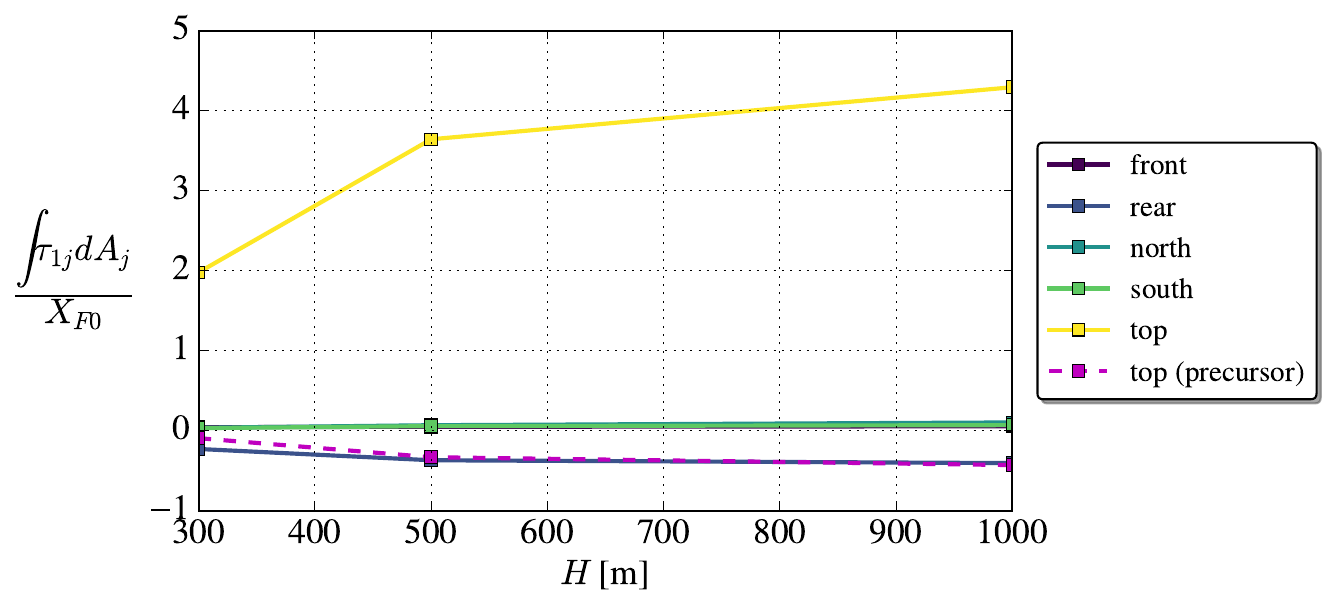}}
  \caption{Contributions to $\Delta M_{\rm{turb}}$ evaluated from LES data at the various surfaces of the CV.}
\label{fig:stress_parts}
\end{figure}

The last contributing term to the momentum availability factor is from turbulence, $\Delta M_{\rm turb}$, which entrains mean momentum into the wind farm through enhanced mixing. It is defined in (\ref{eq:M_contributions}) and it should be noted that it does not include the wall shear stress. Writing out its terms gives
\begin{align}
    \Delta M_{\rm{turb}} &= \frac{ X_{\rm turb} - X_{\rm turb0}}{X_{F0}} 
    = \frac{ \int_{\Omega_{\rm cv}\setminus  \Omega_w}  \tau_{1j} dA_j - \int_{\Omega_{\rm top}}  \tau_{1j,0} dA_j }{X_{F0}} ,
    \label{eq:Mstr_written_out}
\end{align}
where only the top surface contribution from the precursor case is present, because the front/rear and south/north terms cancel in the precursor due to horizontal homogeneity. The above expression is exact, but to obtain a practical model \citet{kirby_ana2023} proceeded without the front/rear and south/north contributions from the farm case. As shown in figure \ref{fig:stress_parts}, this is a good assumption, except for the rear contribution, which is negative (it entrains momentum out of the CV) with a magnitude of around 10\% of the top surface entrainment (which is positive and entrains momentum into the CV). If one nevertheless neglects the rear contribution, the equation simplifies to
\begin{equation}
    \label{eq:deltaMstr_basic}
    \Delta M_{\rm{turb}} 
    = \frac{\tau_t - \tau_{t0} }{\tau_{w0}} , 
\end{equation}
where we have dropped the 13 subscript and, as described below (\ref{eq:Mstr_model}), denote the surface-averaged shear stress on the CV top surface as $\tau_t$. 

\begin{figure}
  \centerline{\includegraphics[width=1\textwidth]{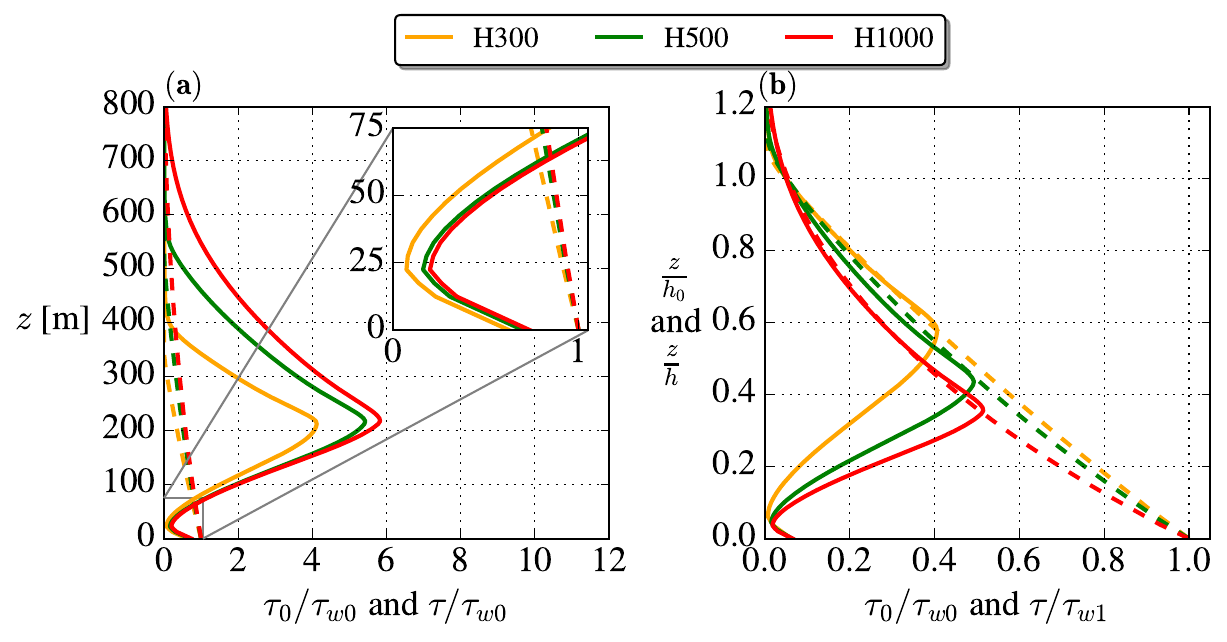}}
  \caption{Streamwise shear stress profiles of the precursor (\dashed) and farm region of the wind farm simulation (\full) (\textbf{a}) and test of their self-similarity (\textbf{b}). The fictive wall shear stress $\tau_{w1} \equiv M \tau_{w0}$ is directly obtained through LES data of $M$ and $\tau_{w0}$. The ABL and farm heights are defined with the 5\% (of $\tau_{w0}$ and $\tau_{w1}$, respectively) rule.}
\label{fig:stress_profile}
\end{figure}

To model $\tau_t$, \citet{kirby_ana2023} assumed that the streamwise shear stress profiles of the ABL and farm simulations, see figure \ref{fig:stress_profile}(a), are self-similar above the top-tip height, i.e.
\begin{equation}
    \frac{\tau_0\left(z/h_0\right)}{\tau_{w0}} = \frac{\tau(z/h)}{\tau_{w1}}  \quad ({\rm for}~z > z_{\rm hub} + D/2),
    \label{eq:self_similarity}
\end{equation}
where a second (fictive) wall shear stress is defined as $\tau_{w1} \equiv M \tau_{w0}$. One can interpret $\tau_{w1}$ as the wall shear stress one would obtain by extrapolating the $\tau(z)$-profile above the top-tip height of the turbines to the ground. Figure \ref{fig:stress_profile}(b) shows that all profiles do not exactly collapse to a single curve, but that each pair of ABL/farm simulations does, which is the important part because this justifies the use of (\ref{eq:self_similarity}). 

Equation (\ref{eq:self_similarity}) can be rearranged to
\begin{align}
    \tau(z) = \tau_{w1}\frac{\tau_0\left(z \frac{h_0}{h}\right)}{\tau_{w0}} ,
\end{align}
which is used to obtain the top shear stress
\begin{align}
    \tau_t = \tau_{w1}\frac{\tau_0\left(H_F \frac{h_0}{h}\right)}{\tau_{w0}} .     \label{eq:taut_eq1}
\end{align}
To use this expression one needs $\tau_0\left(H_F \frac{h_0}{h}\right)$, which \citet{kirby_ana2023} obtained by assuming a linear profile of $\tau_0(z)$ in the interval $z=\left[0,H_F\right]$, i.e.
\begin{align}
    \tau_0(z) = \tau_{w0} + \frac{\tau_{t0} - \tau_{w0}}{H_F} z \quad \quad ({\rm for}~ 0 < z < H_F),
    \label{eq:linear_tau0_assumption}
\end{align}
leading to $\tau_0\left(H_F \frac{h_0}{h}\right) = \tau_{w0} + (\tau_{t0} - \tau_{w0}) \frac{h_0}{h}$. Inserting this expression into (\ref{eq:taut_eq1}) gives
\begin{align}
    \tau_t &= M \left(\tau_{w0} + (\tau_{t0} - \tau_{w0}) \frac{h_0}{h} \right) ,
\end{align}
which combined with (\ref{eq:deltaMstr_basic}) results in \citet{kirby_ana2023}'s model for the turbulence contribution to the momentum availability
\begin{align}
    \Delta M_{\rm{turb}}  = M + M\left(\frac{\tau_{t0}}{\tau_{w0}} - 1\right) \frac{h_0}{h}  - \frac{  \tau_{t0}}{\tau_{w0}} ,\label{eq:Mstr_model_repeated} 
\end{align}
i.e. (\ref{eq:Mstr_model}). Unlike the other sub-models, this expression is implicit ($\Delta M_{\rm{turb}}$ is inside $M$ on the right-hand side), but, due to its simple form, inserting (\ref{eq:Mstr_model_repeated}) into (\ref{eq:M_contributions}) with the other sub-models nevertheless leads to an explicit relation for $M$,
\begin{equation}
    M = \frac{1 + \frac{H_F}{L C_{f0}}  \left( 1 - \beta^2   \right)     - \frac{  \tau_{t0}}{\tau_{w0}}}{ \left(1 -\frac{\tau_{t0}}{\tau_{w0}}\right) \frac{h_0}{h}} .
    \label{eq:M_prereq_to_Mturb}
\end{equation}
This expression is equal to the first version of the KDN-model, $ M_{\rm KDN1}$ (\ref{eq:M_KDN1}), when the BL height model $h_0/h = \beta$ (\ref{eq:bl_height_ratio_model}), is used.

\begin{figure}
  \centerline{\includegraphics[width=1\textwidth]{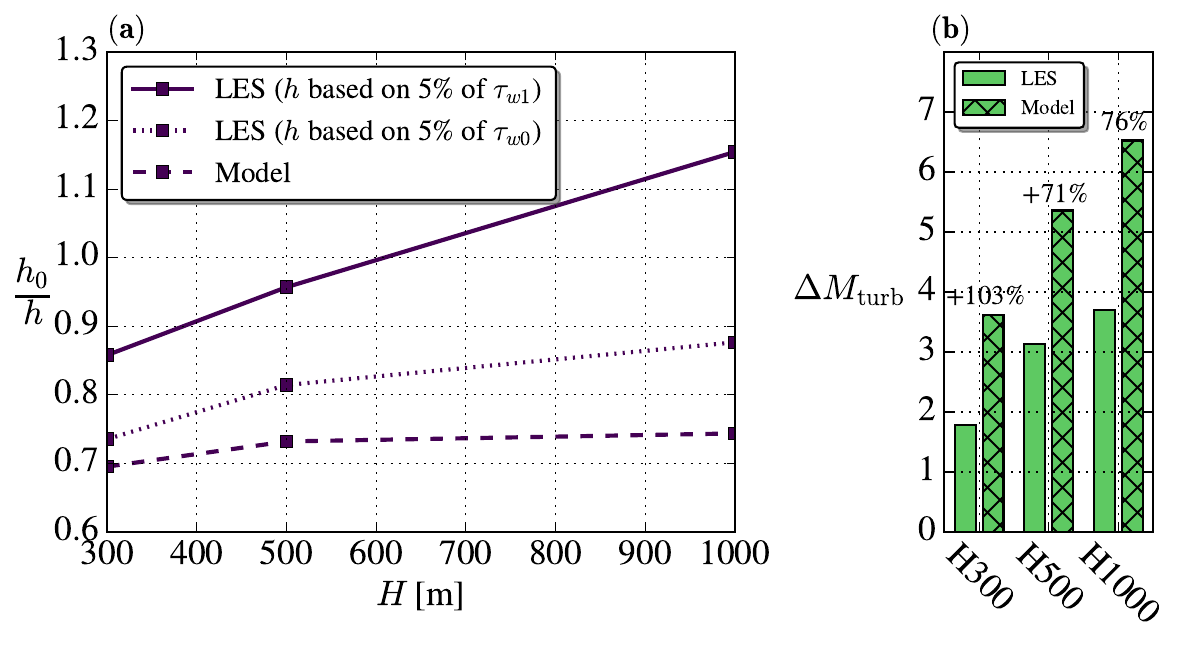}}
  \caption{\textbf{(a)} BL height ratio $h_0/h$ from LES ($h$ based on 5\% of $\tau_{w1}$, c.f. figure \ref{fig:stress_profile}) and model of $h_0/h = \beta$ (\ref{eq:bl_height_ratio_model}). The LES value of $h/h_0$ with $h$ based on 5\% of $\tau_{w0}$, c.f. figure \ref{fig:tau13_contour}, is also displayed, although it is not the relevant height ratio for the $\Delta M_{\rm turb}$ model. \textbf{(b)} Model of turbulence contribution to the momentum availability and its error.}
\label{fig:Mstr_models}
\end{figure}

To test the model of $\Delta M_{\rm turb}$, as shown in figure \ref{fig:Mstr_models}(b), one first evaluates $M$ from (\ref{eq:M_prereq_to_Mturb}) and subsequently inserts this into (\ref{eq:Mstr_model_repeated}). Hence, the turbulence contribution indirectly depends on the other sub-models and the BL height ratio $h_0/h$ through (\ref{eq:M_prereq_to_Mturb}). It should be emphasized that the dependence on the farm BL height $h$ originates from the self-similarity assumption (\ref{eq:self_similarity}), which as shown in figure \ref{fig:stress_profile}(b) is based on the height where the streamwise shear stress is equal to 5\% of $\tau_{w1}$. It thus differs from the farm BL height displayed in figure \ref{fig:tau13_contour}, which was based on 5\% of $\tau_{w0}$, and this explains why, counter-intuitively, $h_0/h$ can exceed one, e.g. the H1000 case in figure \ref{fig:Mstr_models}(a). As discussed in the previous section, the BL height ratio model (\ref{eq:bl_height_ratio_model}) overpredicts $h/h_0$ (based on $\tau_{w0}$) by 6--18\% corresponding to an underprediction of $h_0/h$ of 5--15\%. However, when comparing the model with $h_0/h$ based on $\tau_{w1}$ the underprediction is larger, 19--36\%, which is the main reason for the large overprediction of $\Delta M_{\rm turb}$ shown in figure \ref{fig:Mstr_models}(b). Other sources of errors are from the no rear-entrainment, linear shear stress profile across the CV height and self-similarity assumptions, but these are found to be secondary.

\subsection{Summary of the momentum availability contributions}

\begin{figure}
  \centerline{\includegraphics[width=0.8\textwidth]{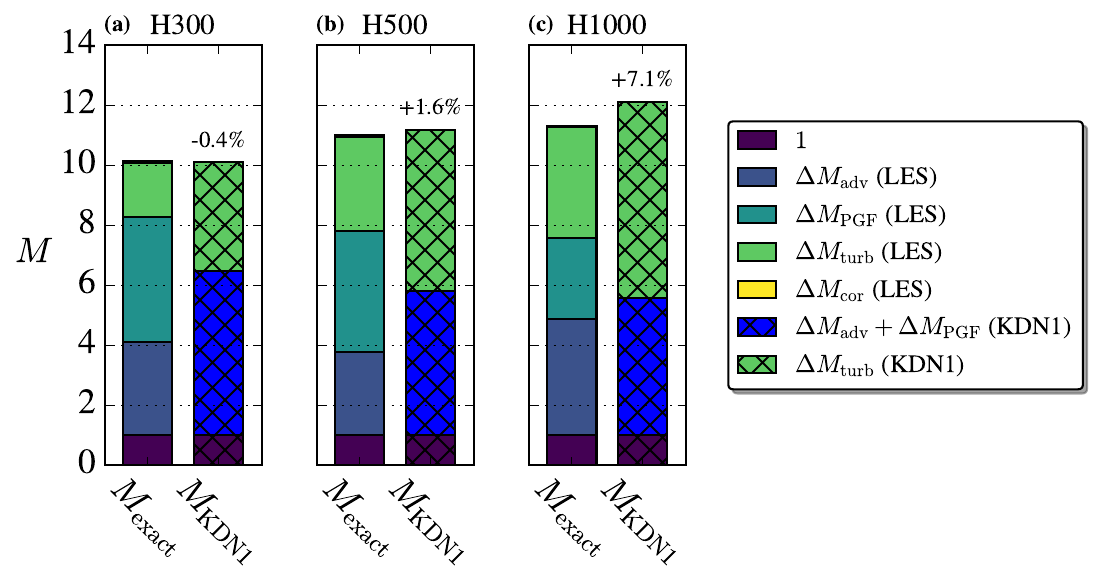}}
  \caption{Split of the momentum availability for the LES data (full) and the $M_{\rm KDN1}$ model (crossed).}
\label{fig:M_split}
\end{figure}

The findings from the previous sections are summarized in figure \ref{fig:M_split}. Each bar has the same total value as the corresponding one from figure \ref{fig:test_of_M}(a-c), but the current figure reveals the split between the various momentum contributions to $M$. From this, we conclude that $\Delta M_{\rm adv}$, $\Delta M_{\rm PGF}$ and $\Delta M_{\rm turb}$ are approximately equally important and that none of these can be neglected. For the $M_{\rm KDN1}$ model, the advection and PGF contributions are grouped together due to the AP approximation (\ref{eq:Madv_and_Mpgf_sa}). The model correctly captures that the turbulence contribution increases with increasing ABL height (this is because the shear stress at the top of the CV increases), although the magnitude is overpredicted as discussed in the previous subsection. Nevertheless, the turbulence overprediction is consistently compensated by the underprediction of the AP approximation, hence the final model prediction of $M_{\rm KDN1}$ is within $-1$\% to 7\% of $M_{\rm exact}$, which is excellent considering the complexity of the momentum availability. Despite the good total performance of $M_{\rm KDN1}$, its practical applicability is limited, since it is non-linear in $\beta$ and a priori requires a value of $\tau_{t0}/\tau_{w0}$, which is the motivation for the $M_{\rm KDN2}$ and $M_{\rm KDN3}$ models.

\subsection{$M_{\rm KDN2}$ and $M_{\rm KDN3}$ models}

The $M_{\rm KDN2}$ and $M_{\rm KDN3}$ models are derived from the $M_{\rm KDN1}$ model as shown in the overview in figure \ref{fig:hierarchy}. $M_{\rm KDN2}$ is first derived by assuming a linear shear stress profile, which is then further simplified to $M_{\rm KDN3}$ by two linearizations for the relationship between $M$ and $\beta$, but the order in which these simplifications are applied is arbitrary and results in the same linear model, $M_{\rm KDN3}$. We have selected the current order of simplifications, because the non-linear $M_{\rm KDN2}$ model is a practical model on its own, for example used by \citet{nishino_power_2025}.

The assumption to simplify $M_{\rm KDN1}$ (\ref{eq:M_KDN1}) to $M_{\rm KDN2}$ (\ref{eq:M_KDN2}) is that $\tau_0(z)$ is linear from $z=0$ to $z=h_0$, which gives
\begin{equation}
    1 -\frac{\tau_{t0}}{\tau_{w0}} \approx \frac{  H_F}{h_0}. 
    \label{eq:linear_shear_stress_assumption}
\end{equation}
This assumption is more restrictive compared to (\ref{eq:linear_tau0_assumption}), where $\tau_0(z)$ was only assumed to be linear up to $z=H_F$. Also, $h_0$ in this context refers not to the 5\% height, but to the full ABL height, i.e. where the streamwise shear stress goes to zero. In the ABL literature \citep[e.g.][]{Berg2020,Liu2021}, $h_0$ is usually defined from the total shear stress $|\tau| = \sqrt{\tau_{13}^2 + \tau_{23}^2}$, which is preferred due to its frame-invariance and because it better represents the actual ABL height, and we therefore choose to use this height as $h_0$ (obtained from LES by fitting $|\tau|_0/|\tau|_{w0} = (1 - z/h_0)^p$, see table \ref{tab:les_precursor_integral_quantities}) in the evaluation of the $M_{\rm KDN2}$ model and in the rest of the paper. As already presented in figure \ref{fig:test_of_M}, the LES data show that the $M_{\rm KDN2}$ model is significantly less accurate than the $M_{\rm KDN1}$ model, especially for taller ABLs, and we will propose a model extension to amend this in §  \ref{sec:rossby}.

\begin{figure}
  \centerline{\includegraphics[width=1\textwidth]{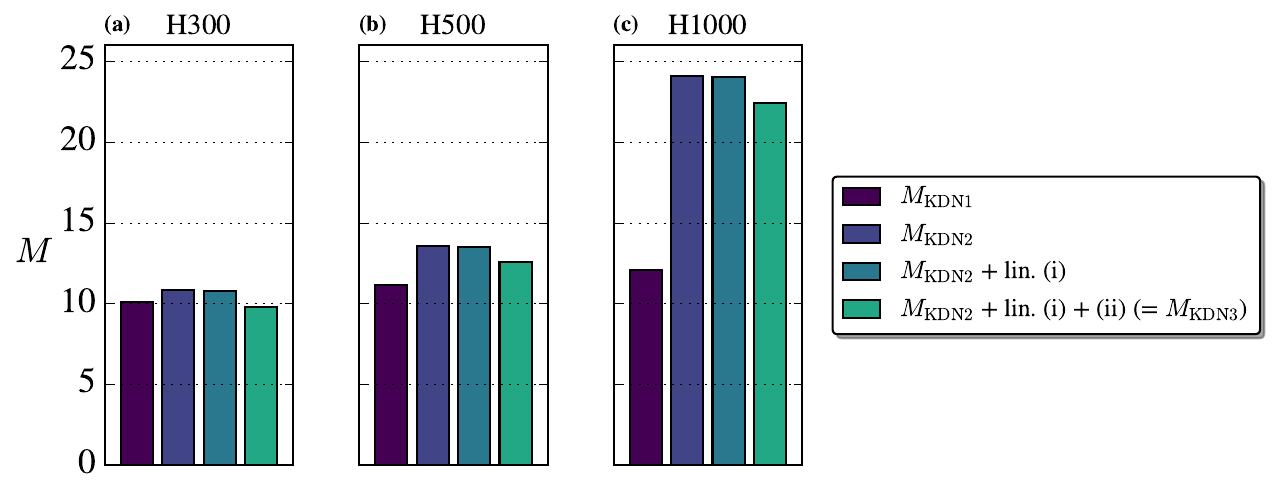}}
  \caption{Test of the two linearizations (\ref{eq:linear_i}-\ref{eq:linear_ii}) used to simplify $M_{\rm KDN2}$ to $M_{\rm KDN3}$. For reference, $M_{\rm KDN1}$ is also included.}
\label{fig:M_appendix_model}
\end{figure}

The $M_{\rm KDN2}$ model (\ref{eq:M_KDN2}) can be rearranged to
\begin{align}
    M_{\rm KDN2}  &=  \frac{1}{1 - (1 - \beta)} +  \frac{ \frac{h_0}{L C_{f0}}  \left( 2\left( 1 - \beta   \right)  - (1 - \beta)^2   \right)      }{1 - (1 - \beta)} ,
    \label{eq:M_KDN_rewritten}
\end{align}
and simplified to $M_{\rm KDN3}$ (\ref{eq:M_KDN3}) using the linearizations suggested by \citet{kirby_ana2023}
\begin{equation}
    \frac{1}{1 - (1 - \beta)} \approx 1 + 1.18 (1 - \beta), \quad \textrm{linearization (i)}, \label{eq:linear_i}
\end{equation}    
\begin{equation}
    \frac{ 2\left( 1 - \beta   \right)  - (1 - \beta)^2   }{1 - (1 - \beta)} \approx 2.18 (1 - \beta) , \quad \textrm{linearization (ii)}.
    \label{eq:linear_ii}
\end{equation}
Figure \ref{fig:M_appendix_model} shows the results of applying these two linearizations. Linearization (i) is applied to the first term in (\ref{eq:M_KDN_rewritten}), which is approximately one order of magnitude smaller than the second term, hence this linearization has little impact on the solution. Adding the second linearization has a larger effect, which, actually, improves the prediction of $M$ slightly, making it closer to $M_{\rm KDN1}$. Overall, both linearizations, however, have a relatively small effect, compared to the linear shear stress assumption (\ref{eq:linear_shear_stress_assumption}) used for simplifying $M_{\rm KDN1}$ to $M_{\rm KDN2}$, and we will therefore focus on modifying this assumption in the next section.

\section{ABL Rossby number extension of $M_{\rm KDN3}$}
\label{sec:rossby}

The $M_{\rm KDN3}$ model (\ref{eq:M_KDN3}) is easy to use in practical applications as it only requires the ABL height $h_0$, farm length $L$, and friction coefficient $C_{f0}$, but as shown in figure \ref{fig:test_of_M}, and in the previous section, it significantly overpredicts $M$ for wind farms operating in tall ABLs. In this section, we will derive a new momentum availability model $M_{\rm BNK}$, see figure \ref{fig:hierarchy}, by including a dependence on the ABL Rossby number $Ro_{h0} \equiv G/(f_c h_0)$. The new model reduces to $M_{\rm KDN3}$ for the special case of $Ro_{h0}^{-1} = 0$ and can thus also be viewed as an extension of this model.

\begin{figure}
    \hspace{-0.5cm}
  \centerline{\includegraphics[width=0.6\textwidth]{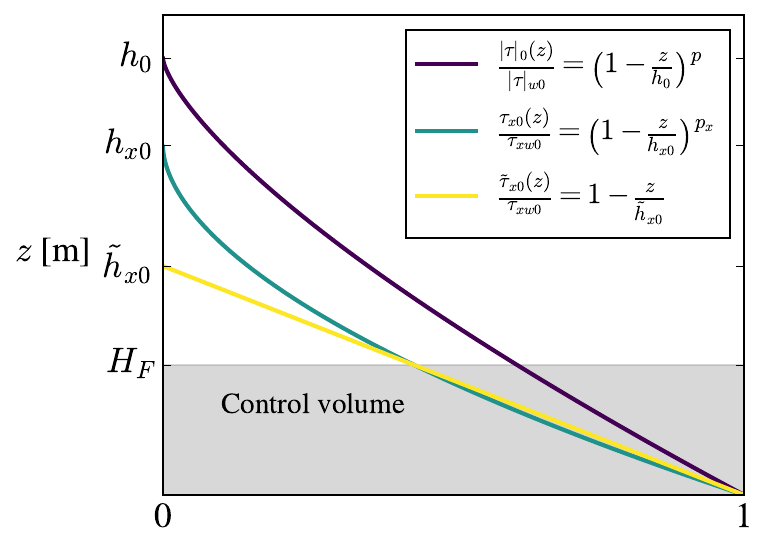}}
  \caption{Sketch of ABL profiles of total shear stress $|\tau|_0(z)$, streamwise shear stress $\tau_{x0}(z)$, and a hypothetical stress $\tilde{\tau}_{x0}(z)$, which is linear and gives $\tilde{\tau}_{x0}(H_F) = \tau_{x0}(H_F) = \tau_{xt0}$.}
\label{fig:sketch_shear_stress_profiles}
\end{figure}

It was identified in the previous section that the principal error of the $M_{\rm KDN3}$ model stems from the assumption 
\begin{equation}
    1 -\frac{\tau_{xt0}}{\tau_{xw0}} \approx \frac{  H_F}{{h}_0},
    \label{eq:main_linear_assumption}
\end{equation}
which comes from assuming a linear profile, $\tau_{x0}(z)/\tau_{xw0} = 1 - z/h_0$ and evaluating it at $z =H_F$ to obtain $\tau_{xt0} = \tau_{x0}(H_F)$. Note that a more explicit notation is used in (\ref{eq:main_linear_assumption}) and the rest of this section, i.e. a subscript $x$ is added to the stresses to denote the direction, compared to in the previous sections (where this subscript was omitted for brevity). The problem with (\ref{eq:main_linear_assumption}) is that $h_0$ is the ABL height defined based on the total shear stress, which is typically larger than the ABL height defined based on the streamwise shear stress, $h_{x0}$. Secondly, the streamwise shear stress profile is not linear, but slightly concave, which implies that an even smaller height, $\tilde{h}_{x0} < h_{x0}$, should be used as sketched in figure \ref{fig:sketch_shear_stress_profiles}. We thus propose to replace (\ref{eq:main_linear_assumption}) with
\begin{equation}
    1 -\frac{\tau_{xt0}}{\tau_{xw0}} \approx \frac{  H_F}{\tilde{h}_{x0}} ,
    \label{eq:main_linear_assumption_new}
\end{equation}
corresponding to assuming a hypothetical linear profile $\tilde{\tau}_{x0}(z)/\tau_{xw0} = 1 - z/\tilde{h}_{x0}$ that gives $\tilde{\tau}_{x0}(H_F) = \tau_{xt0}$. To use (\ref{eq:main_linear_assumption_new}) a model for $\tilde{h}_{x0}$ is needed.

If the streamwise shear stress profile can be expressed as a power function
\begin{equation}
    \frac{\tau_{x0}(z)}{\tau_{xw0}} = \left(1 - \frac{z}{h_{x0}}\right)^{p_x},
    \label{eq:power_law_taux}
\end{equation}
it is simple mathematical problem to determine $\tilde{h}_{x0}$ from $(H_F,h_{x0},p_x)$. Evaluating (\ref{eq:power_law_taux}) at $z = H_F$ to obtain $\tau_{xt0}/\tau_{xw0}$ and inserting this into (\ref{eq:main_linear_assumption_new}) yields
\begin{align}
    \frac{\tilde{h}_{x0}}{H_F}  = \left(1 - \left(1 - \frac{H_F}{h_{x0}} \right)^{p_x}\right)^{-1} .
\end{align}
This expression can with good accuracy be linearized to
\begin{align}
    \frac{\tilde{h}_{x0}}{H_F}  \approx 1 + p_x^{-1.25} \left( \frac{h_{x0}}{H_F} - 1 \right),
\end{align}
for relevant values of $p_x$ and $h_{x0}/H_F$ (for $1 < p_x < 2$ and $1 < h_{x0}/H_F < 5$, the error is around $\pm 3\%$). 

\begin{figure}
  \centerline{\includegraphics[width=1.0\textwidth]{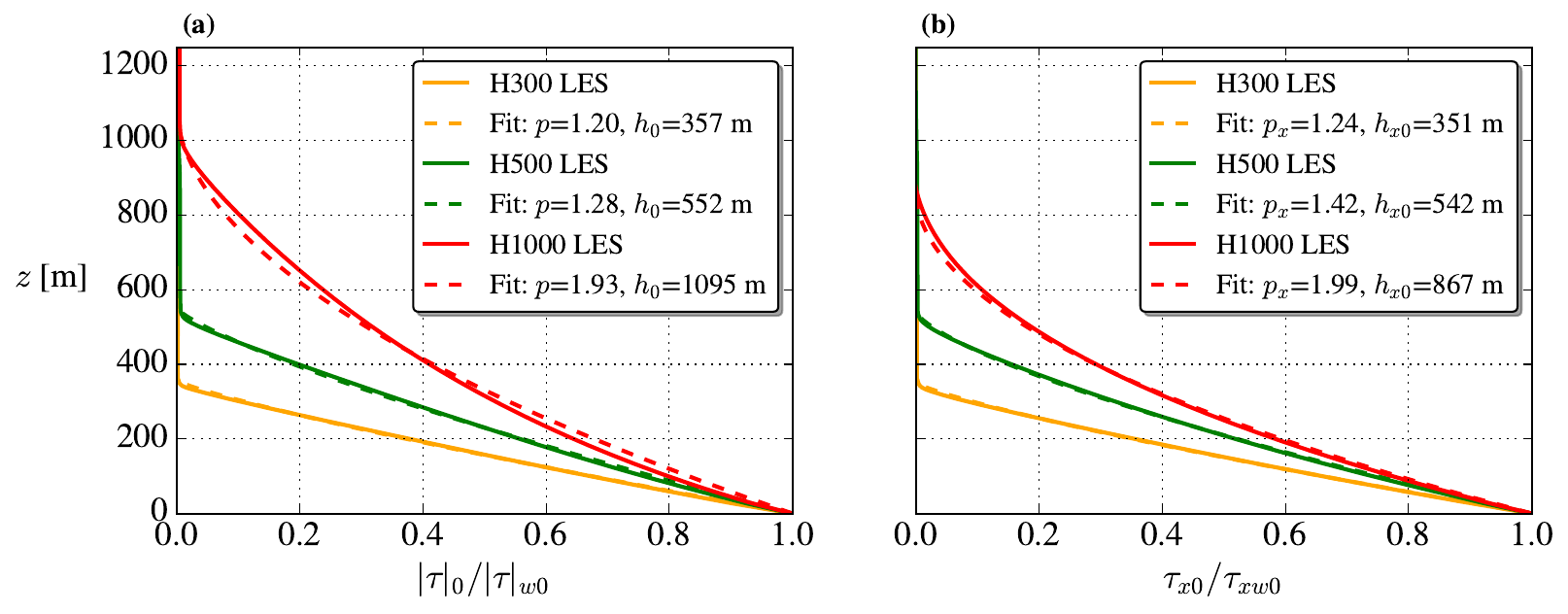}}
  \caption{Power function fits of the \textbf{(a)} total and \textbf{(b)} streamwise shear stress ABL profiles.}
\label{fig:power_law_fits}
\end{figure}

Figure \ref{fig:power_law_fits} shows power function fits to the three ABL profiles of $|\tau|_0(z)$ and $\tau_{x0}(z)$. \citet{nieuwstadt1984} suggested an exponent of $p = 3/2$ for the total shear stress in the stable ABL, which in later works has also been used for the CNBL \citep[e.g.][]{Liu2021,kirby_ana2023,baungaard2024_cnbl}, but the fits in figure \ref{fig:power_law_fits}(a) show that this does not seem to be a universal value for the CNBL. Instead, it appears that the shear stress profile becomes more concave for taller ABLs. Secondly, it is observed that $h_{x0} < h_0$ and $p_x > p$ with the difference stemming from the veering of the shear stress vector caused by the Coriolis effect (in the special case of no Coriolis, $f_c=0$, they would be identical). Thus, even though it was found in § \ref{sec:coriolis} that the Coriolis force itself does not contribute significantly to the momentum availability, i.e. $\Delta M_{\rm Cor} \approx 0$, the Coriolis effect does have an indirect influence on $M$ through $\Delta M_{\rm turb}$ from shear stress veering.

\begin{figure}
  \centerline{\includegraphics[width=1.0\textwidth]{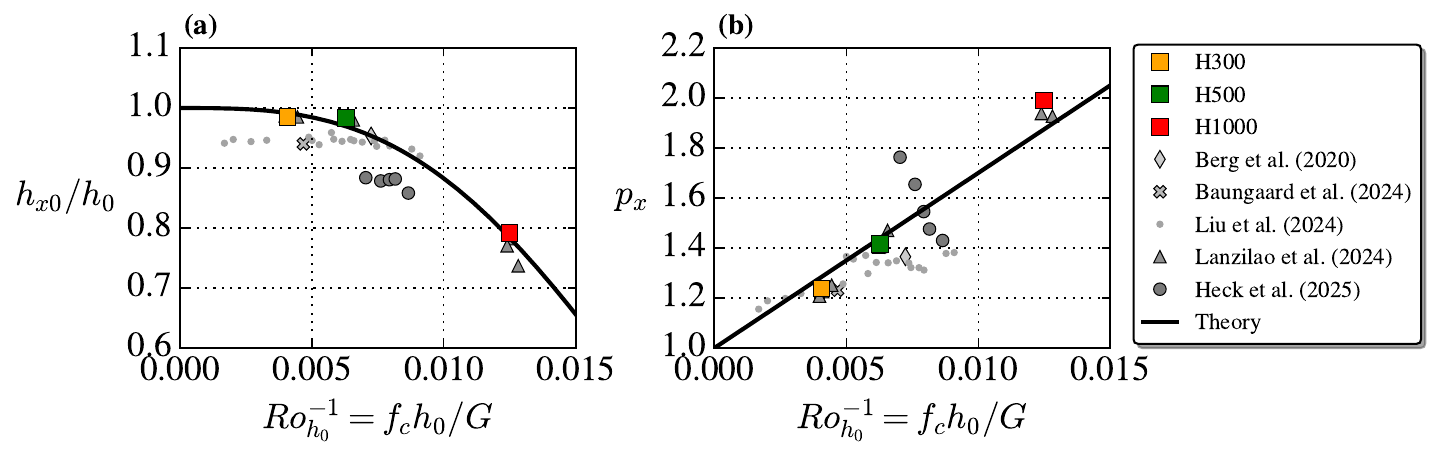}}
  \caption{Fitted parameters from power function fits in figure \ref{fig:power_law_fits}. Additional data from other CNBL LESs \citep{Berg2020,baungaard2024_cnbl,Liu2024,lanzilao2024,Heck2025} are also included.}
\label{fig:h_p_values}
\end{figure}

The fitted values of $h_{x0}/h_0$ and $p_x$ are plotted against the inverse ABL Rossby number, $Ro_{h0}^{-1}$, in figure \ref{fig:h_p_values} along with fits from other CNBL LESs from the literature (all data were rotated to align the hub-height wind direction with the $x$-axis before fitting). Despite some scatter of the data, which can be expected due to different numerics, case setup, integration time, etc., some clear trends of $h_{x0}/h_0$ and $p_x$ as function of $Ro_{h0}^{-1}$ can be observed. It has indeed also been found by \citet{VanderLaan2020} that a Rossby number related to the ABL height (they use a turbulence length scale, $\ell_{\rm max}$, which correlates with the ABL height) is a central parameter for capped ABLs. In the limit of zero Coriolis, $Ro_{h0}^{-1} = 0$, the data are seen to asymptote towards $h_{x0}/h_0 = 1$ and $p_x = 1$ as expected. 

To analytically predict $h_{x0}/h_0$ and $p_x$, we propose some simple theoretical models of these, a super-Gaussian and a linear function, respectively, that satisfy the asymptotic behavior of $h_{x0}/h_0$ and $p_x$ for $Ro_{h0}^{-1} = 0$, as follows

\begin{align}
    \frac{h_{x0}}{h_0} = \textrm{exp}\left( - \left(\frac{Ro_{h0}^{-1}}{0.02}\right)^3 \right),
\end{align}
\begin{align}
    p_x = 1 + 70 Ro_{h0}^{-1} ,
\end{align}
which are plotted together with the data in figure \ref{fig:h_p_values}.

To summarize, the new proposed model is
\begin{equation}
    M_{\rm BNK}  = 1 + \underbrace{\left(1.18  +   2.18 \frac{\tilde{h}_{x0}}{L C_{f0}}    \right)}_{\zeta_{\rm BNK}} (1 - \beta) ,  
    \label{eq:M_new}
\end{equation}
with
\begin{equation}
    \tilde{h}_{x0} = H_F + \left(1 + 70 Ro_{h0}^{-1}\right)^{-1.25} \left( h_{0} \, \textrm{exp}\left( - \left(\frac{Ro_{h0}^{-1}}{0.02}\right)^3 \right) - H_F \right) .
    \label{eq:hx0tilde_model}
\end{equation}
Note that this new model has two additional input parameters, $Ro_{h0}$ and $H_F$, compared to the $M_{\rm KDN3}$ model (\ref{eq:M_KDN3}), but is still posed in a similar linear form for practicality. For the special case of $Ro_{h0}^{-1} = 0$, one obtains $\tilde{h}_{x0} = h_0$ from (\ref{eq:hx0tilde_model}) and the new model is then identical to the $M_{\rm KDN3}$ model. There was no dependence on $H_F$ in the $M_{\rm KDN3}$ model, but since $M$ is independent of $H_F$, whereas $\beta$ is dependent on $H_F$, it is clear from (\ref{eq:M_linear}) that $\zeta$ should, in fact, depend on $H_F$. The inclusion of $H_F$ as a parameter in $\zeta_{\rm BNK}$ therefore has a clear motivation, although it in practice only has a small effect on $\zeta$ for typical values of $H_F$.

\begin{figure}
  \centerline{\includegraphics[width=1\textwidth]{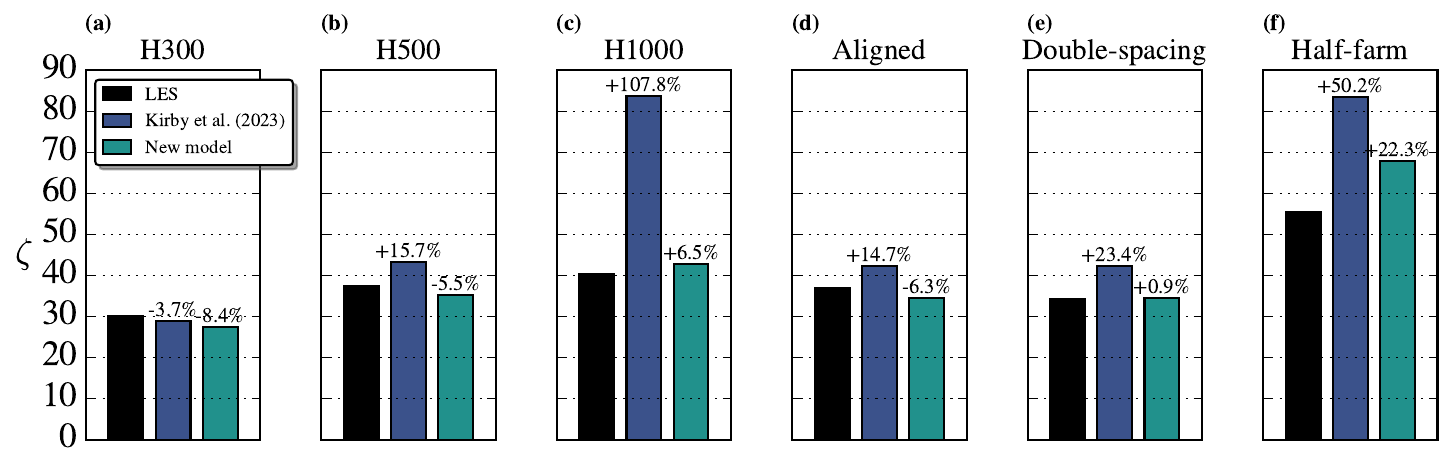}}
  \caption{Momentum response factor from the $M_{\rm KDN3}$ model (\ref{eq:M_KDN3}) and the $M_{\rm BNK}$ model (\ref{eq:M_new}).}
\label{fig:new_model_test_zeta}
\end{figure}

To test the new model, we first compare the momentum response factor
\begin{equation}
    \zeta \equiv \frac{M - 1}{1 - \beta},
\end{equation}
which highlights the principal difference between the proposed $M_{\rm BNK}$ model and the $M_{\rm KDN3}$ model from \citet{kirby_ana2023}. Figure \ref{fig:new_model_test_zeta} shows the comparison, where it is most notable that the new model corrects the previously heavily overpredicted $\zeta$ for the H1000 case. Overall, the new model improves the prediction of $\zeta$ for five out of six cases.

\begin{figure}
\hspace{-0.3cm}
  \centerline{\includegraphics[width=1.05\textwidth]{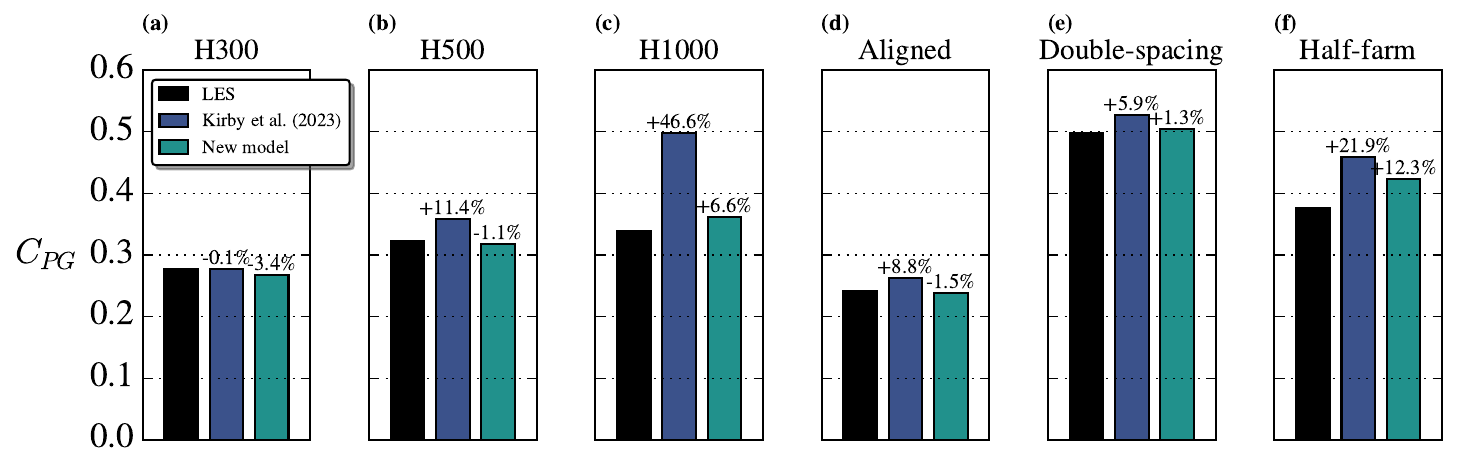}}
  \caption{Global wind farm power efficiency from the $M_{\rm KDN3}$ model (\ref{eq:M_KDN3}) and the $M_{\rm BNK}$ model (\ref{eq:M_new}).}
\label{fig:new_model_test_cpg}
\end{figure}

As a second test, we solve the NDFM equation (\ref{eq:NDFM}), which for linear momentum availability models can be rewritten as
\begin{equation}
    \left(C_{T}^* \frac{\lambda}{C_{f0}} + 1\right) \beta^2 = 1 + \zeta (1 - \beta) ,
    \label{eq:NDFM_linear}
\end{equation}
where we have further assumed $\gamma = 2$ (the actual $\gamma$ values are given in table \ref{tab:les_main_integral_quantities}) to simplify the left-hand side. This latter assumption only introduces a small error, because the first term in (\ref{eq:NDFM}) is dominating as shown earlier in figure \ref{fig:momentum_sinks}, and turns the equation into a quadratic polynomial, which can be solved analytically for $\beta$ (the positive root). With the value of $\beta$ obtained, one can finally predict the global wind farm power efficiency
\begin{equation}
    C_{PG} \equiv \frac{\sum_{i=1}^{N_t} P_i}{\frac{1}{2} \rho N_t A_d U_{F0}^3} = \beta^3 C_P^*,
\end{equation}
where $C_P^* = \frac{\sum_{i=1}^{N_t} P_i}{\frac{1}{2} \rho N_t A_d U_{F}^3}$ is the internal wind farm power coefficient. Figure \ref{fig:new_model_test_cpg} shows the comparison of $C_{PG}$, where an improved prediction is also observed for five out of the six cases with the new model. Again, the most significant improvement is for the tall ABL case, H1000, which indeed also is the case with the largest $Ro_{h0}^{-1}$ value and thus is most affected by the new Rossby sensitization.

\section{Conclusion}
\label{sec:conclusions}

The thrust from wind turbines and the ground (or sea-surface) friction drag are resistances to the atmospheric flow through a wind farm region, and these momentum sinks are balanced by a collection of momentum fluxes from various mechanisms (advection, pressure gradient, Coriolis, turbulence and unsteadiness), which together are called the momentum availability. The momentum availability is a key parameter in the two-scale momentum theory of wind farm flows \citep{Nishino2020}, which is a theoretical modelling framework for predicting wind farm efficiency based on separating turbine/array- and farm/ABL-scales. In this study, we have validated the analytic momentum availability model from \citet{kirby_ana2023} using a LES dataset of wind farm simulations by \citet{lanzilao2024,lanzilao2025} and proposed a new practical extension to improve its generality/accuracy. 

There are three versions of \citet{kirby_ana2023}'s analytic momentum availability model, which we label $M_{\rm KDN1}$, $M_{\rm KDN2}$ and $M_{\rm KDN3}$ in order of their fidelity/complexity, and test against LES data from six wind farm cases. The six cases consist of three with the same turbine layout, but different ABL heights, and three with the same ABL height, but different turbine layouts. In comparison with $M_{\rm exact}$, the exact value from the LES data, the $M_{\rm KDN1}$ model has an error between $-3$\% to 7\% for five out of the six cases, which is excellent given the complexity of the momentum availability. The more practical $M_{\rm KDN2}$ and $M_{\rm KDN3}$ models, which are simpler and require fewer input parameters, on the other hand only agree well with $M_{\rm exact}$ for the lowest ABL case and have significantly larger errors with up to around 100\% overprediction for the tallest ABL case. In general, we observe that the accuracy of all three versions decreases with increasing ABL heights, and we thus focus on the three LES cases with different ABL heights in the majority of the paper.

To understand the validity of the underlying assumptions of the $M_{\rm KDN}$ models, each individual contribution to the momentum availability is evaluated from the LES data, which reveals that the advection, pressure gradient and turbulence terms are the main contributors and of similar magnitude, whereas the contributions from the Coriolis force and unsteadiness can safely be neglected. The advection sub-model of \citet{kirby_ana2023} is found to generally perform well with only small errors coming from the neglected advection out of the lateral sides of the farm and the modelled advection out through the top of the farm. There is a larger error for the pressure gradient sub-model, especially for the case with a low ABL height (H300), but the model nevertheless correctly captures that its contribution decreases with increasing ABL height. This phenomenon can be attributed to the induced favorable pressure gradient through the farm region, which gets stronger as the ABL height is decreased \citep{lanzilao2024}. Finally, the turbulence sub-model requires the most modelling and it is found to be especially sensitive to the modelling of the farm BL height. Overall, however, the sub-models consistently compensate the errors of each other leading to the good overall performance of the $M_{\rm KDN1}$ model. In this analysis, it is also identified that the assumption of a linear shear stress profile in the ABL is the root cause of the degradation of accuracy in the $M_{\rm KDN2}$ model, whereas the two linearizations to obtain $M_{\rm KDN3}$ are fairly accurate.  

A comparison of ABL shear stress profiles from a range of different CNBL studies \citep{Berg2020,baungaard2024_cnbl,Liu2024,lanzilao2024,Heck2025} shows that the concavity $p$ and $p_x$, and ABL height $h_0$ and $h_{x0}$, of the total and streamwise shear stress profiles, respectively, vary considerably. Clear trends of $h_{x0}/h_0$ and $p_x$ as a function of the inverse ABL Rossby number $Ro_{h0}^{-1} = f_c h_0/G$ are identified, for which some simple theoretical models are proposed. Using these models, a non-linear streamwise stress profile can be predicted analytically, which is the basis for our new momentum availability model $M_{\rm BNK}$. This model reduces to $M_{\rm KDN3}$ for the special case of $Ro_{h0}^{-1} = 0$ (thus, one can also view it as an extension of $M_{\rm KDN3}$) and is similarly in a linear form, which makes the model simple and practical to use in wind farm power prediction. The $M_{\rm BNK}$ model amends the issues with the $M_{\rm KDN3}$ model for tall ABL cases. In particular, the error of the two-scale momentum theory prediction of global wind farm power efficiency is reduced from 47\% to 7\% for the case with the tallest ABL height (H1000), while still retaining, or even further improving, its prediction also for lower ABL heights. The $M_{\rm BNK}$ model depends on the Coriolis frequency $f_c$ through $Ro_{h0}^{-1}$ and it is thus expected to be more generally applicable compared to the $M_{\rm KDN3}$ model, while still remaining in a simple linear form, which justifies its position in the hierarchy of momentum availability models in figure \ref{fig:hierarchy}. Its development is based on CNBLs, so its validity in other types of ABLs is not clear, which may limit its application range and could be a topic for future studies. Another interesting future subject is the modelling of momentum availability for wind farm wakes, which is increasingly becoming more important due to the ever growing number of offshore wind farms.


\backsection[Acknowledgements]{M.B. acknowledges the Carlsberg Foundation for his funding, Linacre College for his fellowship and the Department of Engineering Science at the University of Oxford for hosting him. We thank Luca Lanzilao and Johan Meyers for open-sourcing the data from their wind farm LESs, which have been instrumental for this paper. We also thank Kirby Heck, Luoqin Liu and their co-authors for sharing relevant CNBL LES data.}

\backsection[Funding]{This research has been partly supported by the Carlsberg Foundation (grant no. CF-1002) and the UK Natural Environment Research Council (NERC ECOFlow programme, NE/Z504099/1).}

\backsection[Declaration of interests]{The authors report no conflict of interest.}

\backsection[Data availability statement]{The LES data that support the findings of this study are openly available in the KU Leuven RDR database at \href{https://doi.org/10.48804/LRSENQ}{https://doi.org/10.48804/LRSENQ}, \citep{lanzilao2025}. Post-processed data and plotting scripts are available at \url{https://github.com/mchba/Validation-and-extension-of-an-analytic-momentum-availability-model}.} 

\backsection[Author ORCIDs]{M. Baungaard, \url{https://orcid.org/0000-0001-6237-3273}; T. Nishino, \url{https://orcid.org/0000-0001-6306-7702}; A. Kirby, \url{https://orcid.org/0000-0001-8389-1619}}

\backsection[Author contributions]{M.B. post-processed the LES data and produced the figures. T.N. supervised the research. M.B. proposed the ABL Rossby number extension, which builds upon the analytic momentum availability model originally proposed by A.K and T.N. M.B. wrote the paper with corrections from T.N. and A.K.}

\bibliographystyle{jfm}
\bibliography{references}

\end{document}